\documentclass[11pt,a4paper]{article}
\usepackage{jheppub}
\usepackage{subcaption}
\usepackage{multirow}
\usepackage{booktabs}
\usepackage{placeins}
\title{Characteristic Modes of Non-minimally Coupled Thick Branes}

\author{Jing-Xin Gong,}
\author[1]{Heng Guo,\note{Corresponding author.}}
\author{Yong-Tao Lu}
\author{and Xin-Yi Pan}

\affiliation{School of Physics, Xidian University, Xi'an 710071, China}

\emailAdd{24201214709@stu.xidian.edu.cn}
\emailAdd{hguo@xidian.edu.cn}
\emailAdd{luyt@stu.xidian.edu.cn}
\emailAdd{23009200092@stu.xidian.edu.cn}

\abstract{
	Characteristic modes provide important information about the intrinsic properties
	and dynamical response of gravitational systems. In this work, we investigate
	characteristic modes of thick branes with a non-minimally coupled bulk scalar field.
	Two representative thick brane solutions are considered, corresponding to a single-kink scalar configuration and a deformed kink configuration with internal structure, respectively.
	By analyzing the tensor perturbation equations, we first examine the localization of the graviton zero mode in the two brane models.
	For both brane models, the graviton zero mode is localized on the brane.
	No pronounced resonant  Kaluza--Klein (KK) modes are found for the
	single-kink brane at $\xi=1/7$.
For the deformed brane, we investigate the resonant KK modes at
$\xi=0.99$, finding that increasing the deformation parameter $k$
leads to more resonant modes and a longer-lived first resonance.
 We then calculate the quasinormal modes (QNMs) of both brane models independently in the frequency and time domains. The  quasinormal frequencies (QNFs)  obtained by the independent methods agree well for the low overtones.  The masses and decay rates of the resonant KK modes agree 
closely with the real parts and the magnitudes of the imaginary parts of 
the corresponding QNFs, respectively. 
}

\keywords{
	Thick branes, Non-minimal coupling, Resonant Kaluza--Klein modes, Quasinormal modes
}

\begin{document}
\maketitle
\section{Introduction}

Characteristic modes play an important role in characterizing the intrinsic
properties and dynamical response of a physical system. For dissipative
gravitational systems, quasinormal modes (QNMs) describe the response to perturbations and have been extensively studied
in black holes and gravitational systems with higher dimensions
\cite{Kokkotas_1999,Nollert_1999,Berti_2009,Konoplya_2011,Konoplya:2008yy}.
Quasinormal modes have also been investigated in connection with other physical properties
of black holes, including greybody factors, thermodynamic properties, and
black hole shadows~\cite{Lin_2024,Pedrotti_2024}.
In higher-dimensional and braneworld scenarios, QNMs and the associated
ringing phenomena have attracted considerable attention as probes of the
underlying geometry. For example, the QNMs of
brane-localized fields have been studied for higher-dimensional black hole
backgrounds, while characteristic ringing and echoes have also been considered
in braneworld geometries
\cite{Kanti:2005xa,Kanti:2006ua,Bronnikov:2019sbx}. More recently, the spectral stability and time-domain ringdown of thick brane
QNMs under localized deformations of the effective potential have also been
investigated~\cite{Jia_2026}. Scalar-gravitational
quasinormal modes and gravitational echoes associated with the internal
structure of five-dimensional (5D) thick branes have also been studied
in Ref.~\cite{Deng_2026}.
Gravitational quasinormal and metastable modes associated with the braneworld
itself have been investigated in
Refs.~\cite{Seahra_2005_ringing,Seahra_2005_metastable,Tan_2022_QNM,Tan_2024,Tan_2024_Rastall,Jia_2024_finite}.
Together with resonant states, QNMs therefore provide useful probes of the dynamical properties of thick branes.

To place these characteristic modes in their broader theoretical context, we briefly recall the braneworld framework, in which our four-dimensional (4D) universe is embedded in a higher-dimensional spacetime. Early
brane and domain-wall scenarios suggested that ordinary matter may be confined
to a lower-dimensional hypersurface embedded in a higher-dimensional
spacetime
\cite{Akama_1982,Rubakov_1983,Visser_1985,Antoniadis_1990}.
In the large extra dimension scenario, the weakness of gravity can be
attributed to the large volume of the compact extra dimensions
\cite{ArkaniHamed_1998,Antoniadis_1998,ArkaniHamed_1999}.
The Randall--Sundrum models subsequently showed that a warped extra dimension
can provide an exponential mechanism for generating the hierarchy between the
Planck and electroweak scales
\cite{Randall_1999a,Randall_1999b}. In the single brane model, the graviton
zero mode is localized near the brane, allowing effective 4D gravity to be
recovered. These warped braneworld scenarios have since been investigated
extensively in gravitational, cosmological, and field theoretical contexts
\cite{Chamblin_2000,Shiromizu_2000,Garriga_2000,Giddings_2000,Karch_2001,
	Csaki_2004,Rubakov_2001,Brax_2004,Maartens_2010}.

Since the original Randall--Sundrum branes are idealized as infinitely thin
objects, smooth thick brane configurations generated dynamically by background
scalar fields were subsequently introduced
\cite{DeWolfe_2000,Gremm_2000,Csaki_2000,Dzhunushaliev_2010}.
A large variety of thick brane solutions and their extensions have been
constructed in different gravitational and matter fields 
\cite{Bazeia_2003,Bazeia_2004,Bazeia_2006,Afonso_2006,
	BarbosaCendejas_2005,BarbosaCendejas_2006,Arias_2002,Kobayashi_2002,
	Kehagias_2001,Koley_2005,Dzhunushaliev_2008,
	Bazeia_2009,Liu_2009_split,Bazeia_2014,Correa_2011}.
Compared with thin branes, these smooth configurations provide a natural
framework for studying the internal structure and splitting of the brane,
as well as the localization of gravity and matter fields and the associated Kaluza--Klein (KK) spectra. The localization of scalar, vector, antisymmetric tensor, and
fermion fields has therefore been widely investigated in different
thick brane backgrounds
\cite{Bajc_2000,Oda_2000,Ringeval_2002,Melfo_2003,Dubovsky_2000,
	Liu_2008_scalar,Liu_2008_fermion,Liu_2009_fermion,Almeida_2009,
	Chumbes_2012,Zhao_2010,Zhao_2011,Cruz_2013_Bloch,Guo_2011_KR,
	Landim_2011,Fu_2011,Fu_2012,Xie_2013,
	Liu_2018_review}. In particular, the internal structure of a thick brane can
significantly modify the corresponding effective potentials and hence the
localization and resonance spectra 
\cite{Bazeia_2003,Kehagias_2001,Koley_2005,Liu_2009_split,Correa_2011}.

Scalar fields may couple non-minimally to gravity in scalar--tensor theories.
The Brans--Dicke theory provides one of the earliest examples of a scalar
degree of freedom coupled to gravity
\cite{Brans_1961,Fujii_2003}, while more general scalar--tensor and
modified gravity theories have been developed in a variety of contexts
\cite{Faraoni_2004,Sotiriou_2010,DeFelice_2010,Clifton_2012,
	Nojiri_2011,Nojiri_2017,Capozziello_2011,Olmo_2011}.
In thick brane scenarios, a non-minimal scalar-curvature coupling can be
introduced through a function $F(\phi)$ multiplying the scalar curvature, and
various thick brane solutions with such a coupling have been constructed
~\cite{Bogdanos_2006,Liu_2012,Guo_2012}.
Such a coupling modifies the background geometry and can significantly affect
the internal structure and splitting behavior of the brane, as well as the
localization of gravity.
Related perturbation and localization problems have also been studied in thick
brane models with modified gravitational dynamics
\cite{Zhong_2011_tensor_fR,Zhong_2013_Kbrane,Guo_2025}.

To study the localization of gravity and the  associated KK spectrum,
we consider transverse-traceless tensor perturbations, whose perturbation
equation can generally be reduced to a Schr\"odinger-like equation.
The corresponding effective potential determines the localization of the
graviton zero mode and the behavior of the massive KK modes
\cite{DeWolfe_2000,Gremm_2000,Csaki_2000,Guo_2012}.
The localized zero mode is essential for recovering effective 4D gravity,
whereas the spectrum of massive KK modes contains additional information about the
extra-dimensional geometry. For a noncompact extra dimension, massive KK
modes are generally not normalizable bound states. Nevertheless, some massive
modes can remain temporarily concentrated near the brane and appear as
resonant KK modes. Such quasi-localized KK modes have been studied in a variety of braneworld configurations
\cite{Csaki_2000_resonantGravity,Liu_2009_resonance,Cruz_2014_gravitonRes,Yu_2016_fRRes,
	Tan_2021_fTRes,Chen_2021_mimeticfR}. Their masses, lifetimes, and resonance
spectra depend sensitively on the structure of the effective potential and
can therefore provide information about the internal structure of the brane.
Since resonant states are metastable excitations, their time evolution can
also be used to determine their characteristic oscillation frequencies and
decay behavior
\cite{Tan_2023_scalarRes}.

This oscillatory decay suggests a natural connection between resonant KK modes and QNMs. Early studies of the Randall--Sundrum braneworld showed that metastable massive gravitational modes can be described as QNMs \cite{Seahra_2005_ringing,Seahra_2005_metastable}, which were subsequently investigated in thick brane systems \cite{Tan_2022_QNM}. More recently, it was shown that resonant KK modes
and QNMs can provide closely related descriptions of the characteristic modes
of a thick brane: the oscillation frequencies of the resonances agree with the
real parts of the corresponding quasinormal frequencies (QNFs), while their
decay rates agree with the magnitudes of the imaginary parts
\cite{Tan_2024}. Quasinormal modes and related characteristic modes have also been investigated in thick brane systems with modified gravitational dynamics and nontrivial internal structures
\cite{Tan_2025_graviscalar,YuPeng_2025_fRQNM,Pan_2026}.
These developments motivate a further examination of whether the resonance--QNM
correspondence persists in non-minimally coupled thick branes. We also analyze
how the non-minimal coupling and the internal structure of the brane affect the
localization and characteristic modes of tensor perturbations.

In this work, we investigate the characteristic gravitational modes of two representative thick brane solutions with a non-minimally coupled bulk scalar field, corresponding to a single-kink scalar configuration and a deformed kink configuration with internal structure, respectively. We first study the localization of the graviton zero mode in both brane models.
We then investigate
the resonant KK modes and QNMs of the two models and examine the
resonance--QNM correspondence.

The paper is organized as follows. In Sec.~\ref{sec:review}, we review the
non-minimally coupled thick brane models and introduce Model A and Model B.
In Sec.~\ref{sec:localization}, we study the localization of the graviton zero mode and the resonant KK modes, with emphasis on the resonance structure and time evolution in Model B. In Sec.~\ref{sec:qnm}, we calculate the QNMs of the
two models and discuss their relation to the resonant KK modes. Finally, the
conclusions and possible extensions are presented in
Sec.~\ref{sec:conclusion}.

\section{Review of non-minimally coupled thick brane models}\label{sec:review}
	\label{BRANE WORLD MODEL}
We study a  5D gravity theory with a non-minimally coupled bulk scalar field. The action is written as
	\begin{equation}
		S=\int d^5x\sqrt{-g}
		\left[
		\frac{1}{2\kappa^2_5}F(\phi)R
		-\frac{1}{2}g^{MN}\partial_M\phi\partial_N\phi
		-V(\phi)
		\right],
		\label{eq:action_nonminimal}
	\end{equation}
where \(R\) is the 5D Ricci scalar, \(\phi\) denotes the
background scalar field, and \(V(\phi)\) is its potential. We choose the
non-minimal coupling function as
\begin{equation}
	F(\phi)=1-\xi\phi^2,
	\label{eq:F_phi_unified}
\end{equation}
where \(\xi\) is the non-minimal coupling parameter. Moreover,
\(\kappa_5^2=8\pi G_5\), where \(G_5\) is the 5D Newton
constant. Throughout this paper, we set \(\kappa_5^2=1\).
	The Einstein equations corresponding to the action \eqref{eq:action_nonminimal}
	are expressed as follows
\begin{equation}
	F(\phi)
	\left(
	R_{MN}-\frac{1}{2}g_{MN}R
	\right)
	+
	\left(
	g_{MN}\Box-\nabla_M\nabla_N
	\right)F(\phi)
	=
	T^{(\phi)}_{MN},
	\label{eq:modified_einstein}
\end{equation}
where $T^{(\phi)}_{MN}
	=
	\partial_M\phi\partial_N\phi
	-g_{MN}
	\left[
	\frac{1}{2}g^{PQ}\partial_P\phi\partial_Q\phi
	+V(\phi)
	\right]$ is the energy-momentum tensor of the scalar field, and $\Box$ is the 5D d'Alembertian operator. The Einstein
	equations can be rewritten in the standard form
	\begin{equation}
		R_{MN}-\frac{1}{2}g_{MN}R
		=
		T^{\rm eff}_{MN},
		\label{eq:einstein_eff}
	\end{equation}
	where the effective energy-momentum tensor is defined as
	\begin{equation}
		T^{\rm eff}_{MN}
		=
		\frac{1}{F(\phi)}
		\left[
		T^{(\phi)}_{MN}
		-\left(g_{MN}\Box-\nabla_M\nabla_N\right)F(\phi)
		\right].
		\label{eq:Teff_nonminimal}
	\end{equation}
The equation of motion for the scalar field is
\begin{equation}
	\Box\phi+\frac{1}{2}F_\phi(\phi)R-V_\phi(\phi)=0,
	\label{eq:scalar_nonminimal}
\end{equation}
where $F_\phi(\phi)\equiv dF/d\phi$ and $V_\phi(\phi)\equiv dV/d\phi$. The scalar equation of motion is not independent of the Einstein
equations~\cite{Farakos_2006_RS2Nonminimal}.
The metric describing a static flat brane is given by
	\begin{equation}
		ds^2=e^{2A(y)}\eta_{\mu\nu}dx^\mu dx^\nu+dy^2,
		\label{eq:metric_y}
	\end{equation}
where $e^{2A(y)}$ is the warp factor, \(\eta_{\mu\nu}=\mathrm{diag}(-1,+1,+1,+1)\) is the Minkowski metric, and $y$ is the extra-dimensional
coordinate.  The background scalar field is assumed to depend only on \(y\), namely, \(\phi=\phi(y)\). For the metric \eqref{eq:metric_y}, the scalar curvature is 
	\begin{equation}
		R(y)=-8A''(y)-20A'^2(y),
		\label{eq:Ricci_y}
	\end{equation}
	where the prime denotes the derivative with respect to \(y\).
	
	Substituting Eq.~\eqref{eq:metric_y} into Eqs.~\eqref{eq:modified_einstein}--\eqref{eq:scalar_nonminimal}, we obtain the following background Einstein equations:
\begin{align}
	3F(\phi)\left(4A'^2+A''\right)
	+7A'F'(\phi)+F''(\phi)
	&=-2V(\phi),
	\label{eq:bkg1}
	\\
	3F(\phi)A''
	-A'F'(\phi)
	+F''(\phi)
	+\phi'^2
	&=0,
	\label{eq:bkg2}
	\\
	\phi''
	+4A'\phi'
	-2F_{\phi}(\phi)
	\left(2A''+5A'^2\right)
	-V_{\phi}(\phi)
	&=0.
	\label{eq:bkg3}
\end{align}
	 We introduce the conformal coordinate \(z\) through the
	transformation
	\begin{equation}
		dz=e^{-A(y)}dy.
	\end{equation}
With \(A(z)\equiv A[y(z)]\), the metric takes the conformally flat form
	\begin{equation}
		ds^2=e^{2A(z)}
		\left(
		\eta_{\mu\nu}dx^\mu dx^\nu+dz^2
		\right).
		\label{eq:metric_z}
	\end{equation}
	For the metric \eqref{eq:metric_z}, the corresponding Ricci scalar is
	\begin{equation}
		R(z)=-e^{-2A(z)}
		\left(
		8A''(z)+12A'(z)^2
		\right).
		\label{eq:ricci_scalar_z}
	\end{equation}
Here the prime denotes differentiation with respect to \(z\).
	
	Substituting the conformally flat metric \eqref{eq:metric_z} into the
	Einstein equations and the scalar field equation, we obtain the following
	background equations:
	\begin{align}
		3F(\phi)\left(A''+3A'^2\right)
		+6A'F'(\phi)
		+F''(\phi)
		+2e^{2A}V(\phi)&=0,
		\label{eq:bg_z_1}\\
		3F(\phi)\left(A''-A'^2\right)
		-2A'F'(\phi)
		+F''(\phi)
		+\phi'^2&=0,
		\label{eq:bg_z_2}\\
		\phi''
		+3A'\phi'
		-2F_\phi(\phi)\left(2A''+3A'^2\right)
		-e^{2A}V_\phi(\phi)&=0.
		\label{eq:bg_z_3}
	\end{align}
We consider two representative thick brane solutions to these field equations, which are examined in detail below.

\subsection{Model A: single-kink thick brane solution}
	\label{subsec:modelA}
\begin{figure}[htbp]
	\centering
	
	\begin{minipage}[b]{0.42\textwidth}
		\centering
		\includegraphics[width=\linewidth]{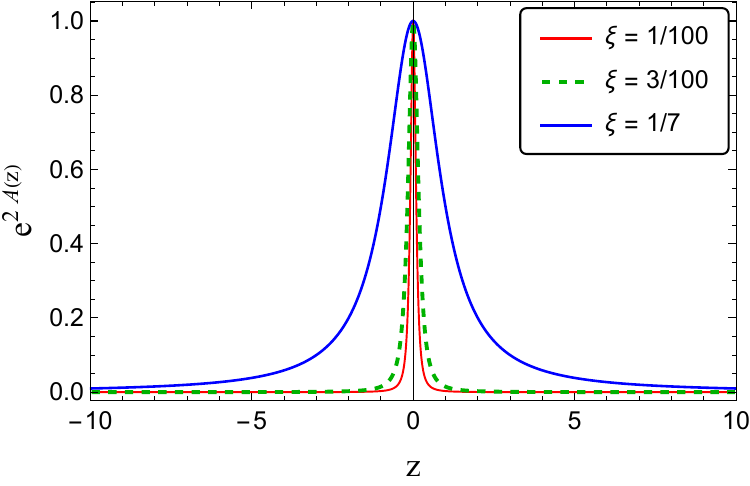}
	\end{minipage}
\hspace{0.04\textwidth}
	\begin{minipage}[b]{0.42\textwidth}
		\centering
		\includegraphics[width=\linewidth]{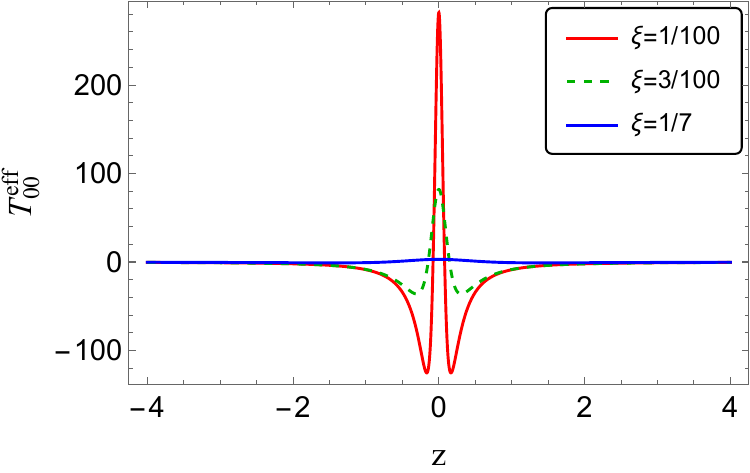}
	\end{minipage}
	
	\vspace{0.2cm}
	
	\begin{minipage}[b]{0.42\textwidth}
		\centering
		\includegraphics[width=\linewidth]{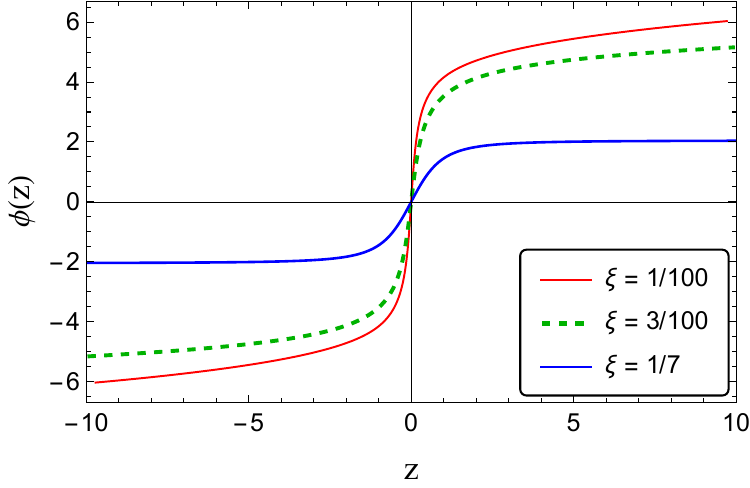}
	\end{minipage}
\hspace{0.04\textwidth}
	\begin{minipage}[b]{0.42\textwidth}
		\centering
		\includegraphics[width=\linewidth]{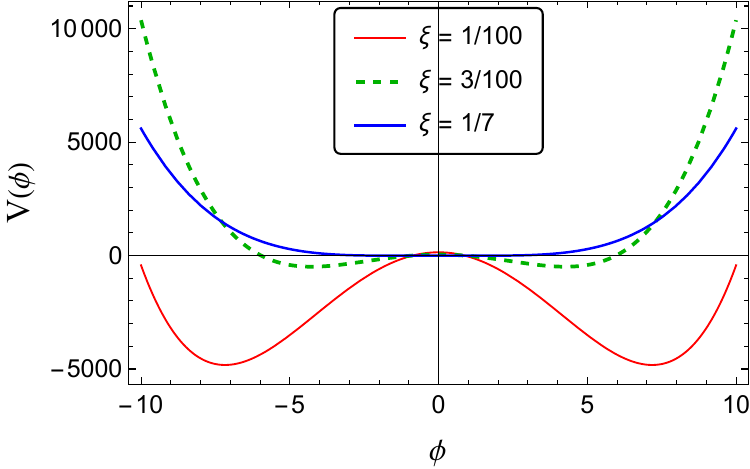}
	\end{minipage}
	
	\caption{
		The profiles of the warp factor \(e^{2A(z)}\), the effective energy density \(T_{00}^{\mathrm{eff}}(z)\), the scalar field \(\phi(z)\),
		and the scalar potential \(V(\phi)\) for Model A.
	}
	
	\label{fig:modelA}
\end{figure}

Model A is a single-kink thick brane solution to the
background equations~\eqref{eq:bkg1}--\eqref{eq:bkg3}, first derived
by Bogdanos, Dimitriadis, and Tamvakis~\cite{Bogdanos_2006} and later
studied in detail in Ref.~\cite{Liu_2012}. In the extra-dimensional
coordinate $y$, the scalar field and the warp factor read
\begin{equation}
	\phi(y)=\phi_0\tanh(\kappa y),
	\qquad
	\phi_0=
	\sqrt{
		\frac{3(1-6\xi)}
		{\xi(1-2\xi)}
	},
	\label{eq:phi_y_modelA}
\end{equation}
and
\begin{equation}
	A(y)=
	\left(6-\frac{1}{\xi}\right)\ln\cosh(\kappa y).
	\label{eq:A_y_modelA}
\end{equation}
Here \(\kappa>0\) sets the characteristic inverse length scale of the scalar field kink and hence controls the brane thickness. For computational convenience, we set \(\kappa=1\) throughout the analysis of Model A.
The corresponding scalar potential is
\begin{equation}
	V(\phi)
	=
	\frac{\kappa^2}{6\xi}
	\left[
	\frac{9-54\xi}{1-2\xi}
	+6\left(\xi(7+24\xi)-2\right)\phi^2
	+\frac{\xi(1-2\xi)(3-16\xi)(4-21\xi)}
	{1-6\xi}\phi^4
	\right].
	\label{eq:V_phi_modelA}
\end{equation}
The effective energy density \(T^{\rm eff}_{00}(y)\) is given by
\begin{equation}
	T^{\rm eff}_{00}(y)
	=
	-3e^{2A(y)}
	\left[
	A''(y)+2A'^2(y)
	\right].
	\label{eq:Teff00_modelA_general}
\end{equation}
This solution is valid for $0<\xi<\frac{1}{6}$.
It is worth noting that, for this 
non-minimally coupled
solution, $V(\phi(y))$ does not approach its vacuum 
values in the limit $y\rightarrow\pm\infty$. Through the coordinate transformation $dz=e^{-A(y)}dy$, we obtain the corresponding numerical solutions in the conformal coordinate $z$. As shown in Fig.~\ref{fig:modelA}, 
the parameter \(\xi\) characterizes the degree of symmetry breaking,
with smaller values of \(\xi\) corresponding to more pronounced
symmetry breaking. However, this non-minimally coupled solution does not admit the limit $\xi=0$. Moreover, the condition \(F(\phi)>0\) is satisfied only for
\(\frac{1}{8}\leq\xi<\frac{1}{6}\).

\subsection{Model B: deformed thick brane solution}
	\label{subsec:modelB}
Model B is a deformed thick brane solution with internal structure. The scalar field  is prescribed in the
conformal coordinate $z$ as
\begin{equation}
	\phi(z)=\phi_{0}\tanh^{k}(bz),	\label{eq:modelB_phi} 
\end{equation}
where $\phi_{0}$ and $b$ are positive constants and the deformation parameter $k$ is a positive
odd integer~\cite{Guo_2012}. As $k$ increases, the scalar field profile becomes progressively flatter around
$z=0$. Therefore, we consider the representative cases
$k = 1, 3, 15$ in this work.

Substituting the coupling function in
Eq.~\eqref{eq:F_phi_unified} into Eq.~\eqref{eq:bg_z_2},
we obtain
\begin{equation}
	3(1-\xi\phi^2)(A''-A'^2)
	+4\xi\phi A'\phi'
	+(1-2\xi)\phi'^2
	-2\xi\phi\phi''=0 .
	\label{eq:modelB_AZ_equation}
\end{equation}
In general, Eq.~\eqref{eq:modelB_AZ_equation} cannot be solved
analytically. Therefore, ${A(z)}$ is obtained
numerically with the boundary conditions
\begin{equation}
	A(0)=0,\qquad A'(0)=0 .
	\label{eq:modelB_boundary}
\end{equation}
To understand the behavior of the warp factor, we expand
the background scalar field around $z=0$. From
Eq.~\eqref{eq:modelB_phi}, one has
\begin{equation}
	\phi(z)
	=\phi_0(bz)^k
	-\frac{k}{3}\phi_0(bz)^{k+2}
	+\mathcal{O}\!\left((bz)^{k+4}\right).
	\label{eq:modelA_phi_expansion}
\end{equation}
In the asymptotic region $z\rightarrow \pm\infty$, one has
$\phi\rightarrow \pm\phi_0$, $\phi'\rightarrow 0$, and
$\phi''\rightarrow 0$. Hence Eq.~\eqref{eq:modelB_AZ_equation}
reduces to
\begin{equation}
	A''(z)-A'^2(z)=0 .
\end{equation}
Following Ref.~\cite{Guo_2012}, the asymptotic form of the warp
factor can be written as
\begin{equation}
	e^{A(z)} = \frac{1}{c+b|z|}
	\simeq 
\frac{1}{b|z|},
	\qquad |z|\rightarrow \infty ,
	\label{eq:modelB_A_asymptotic}
\end{equation}
which has the same asymptotic behavior as the Randall--Sundrum
warp factor.

For $k=1$, the scalar field behaves as
\begin{equation}
	\phi(z)=\phi_0 bz-\frac{1}{3}\phi_0(bz)^3
	+\mathcal{O}\!\left((bz)^5\right).
\end{equation}
Substituting this expansion into
Eq.~\eqref{eq:modelB_AZ_equation}, the warp factor near $z=0$ is
obtained as
\begin{equation}
	A(z)=\frac{1}{6}\phi_0^2(2\xi-1)(bz)^2
	+\mathcal{O}\!\left((bz)^3\right).
	\label{eq:modelB_A_k1}
\end{equation}
Thus, the critical value is $\xi_0=\frac{1}{2}$.
For $\xi<\xi_0$, the warp factor has a maximum at the origin,
corresponding to a single brane. For $\xi>\xi_0$, the origin changes
from a maximum to a minimum and the single brane splits into two
sub-branes.

For $k=3$, the scalar field behaves as
\begin{equation}
	\phi(z)=\phi_0(bz)^3+\mathcal{O}\!\left((bz)^5\right).
\end{equation}
Expanding the warp factor about the symmetry point \(z=0\), we obtain
\begin{equation}
	A(z)=\frac{1}{30}\phi_0^2(10\xi-3)(bz)^6
	+\mathcal{O}\!\left((bz)^7\right),
	\label{eq:modelB_A_k3}
\end{equation}
which gives the critical value $\xi_0=3/10$. The brane remains split into two sub-branes because the scalar field has a double-kink configuration.

\begin{figure*}[t]
	\centering
	
	\begin{subfigure}[b]{0.42\textwidth}
		\centering
		\includegraphics[width=\linewidth]{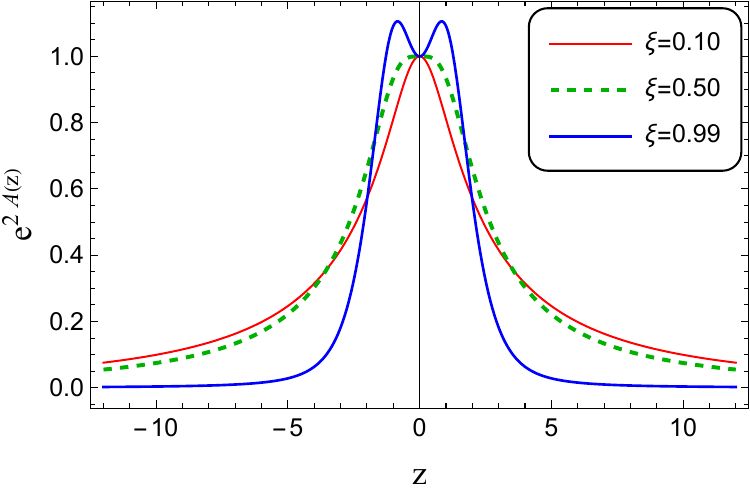}
		\label{fig:k1_warp}
	\end{subfigure}
	\hspace{0.04\textwidth}
	\begin{subfigure}[b]{0.42\textwidth}
		\centering
		\includegraphics[width=\linewidth]{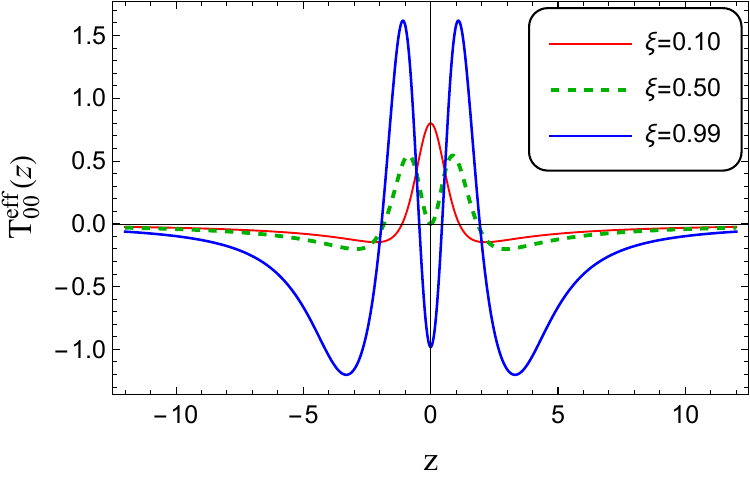}
		\label{fig:k1_T00}
	\end{subfigure}
	
	\vspace{0.2cm}
	
	\begin{subfigure}[b]{0.42\textwidth}
		\centering
		\includegraphics[width=\linewidth]{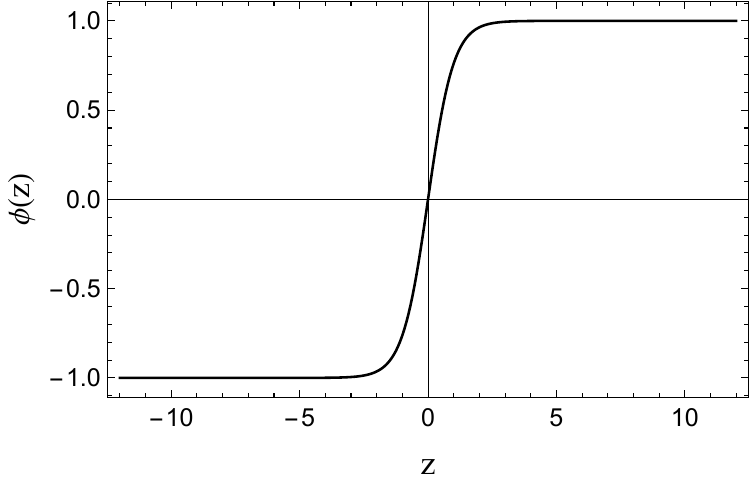}
		\label{fig:k1_phi}
	\end{subfigure}
	\hspace{0.04\textwidth}
	\begin{subfigure}[b]{0.42\textwidth}
		\centering
		\includegraphics[width=\linewidth]{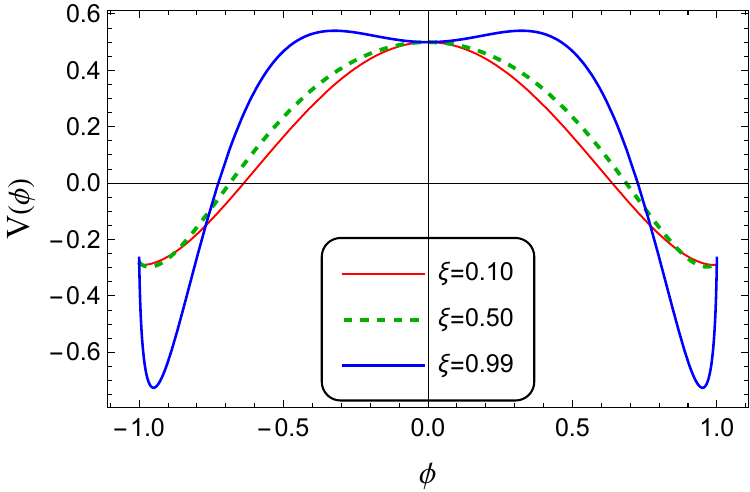}
		\label{fig:k1_Vphi}
	\end{subfigure}
	
	\caption{
		The brane configuration for $k=1$, including the warp factor
		$e^{2A(z)}$, the effective energy density  $T_{00}^{\rm eff}(z)$, the scalar field
		$\phi(z)$, and the scalar potential $V(\phi)$. The other parameters
		are set as $\phi_0=1$ and $b=1$.
	}
	\label{fig:k1_background}
\end{figure*}

\begin{figure*}[t]
	\centering
	
	\begin{subfigure}[b]{0.42\textwidth}
		\centering
		\includegraphics[width=\linewidth]{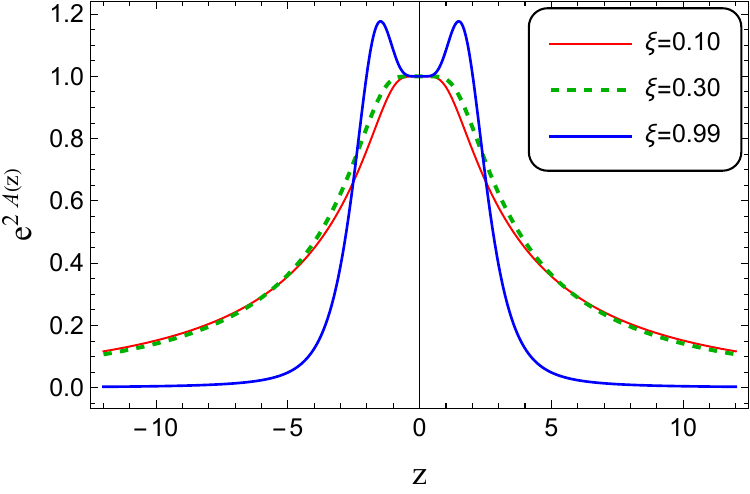}
		\label{fig:k3_warp}
	\end{subfigure}
	\hspace{0.04\textwidth}
	\begin{subfigure}[b]{0.42\textwidth}
		\centering
		\includegraphics[width=\linewidth]{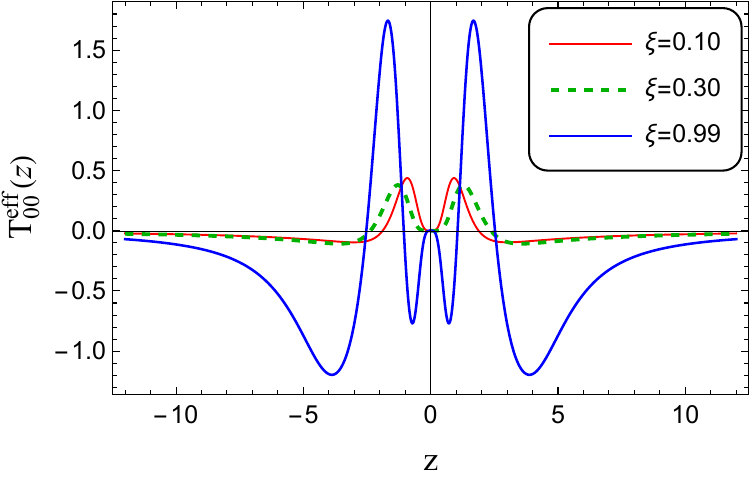}
		\label{fig:k3_T00}
	\end{subfigure}
	
	\vspace{0.2cm}
	
	\begin{subfigure}[b]{0.42\textwidth}
		\centering
		\includegraphics[width=\linewidth]{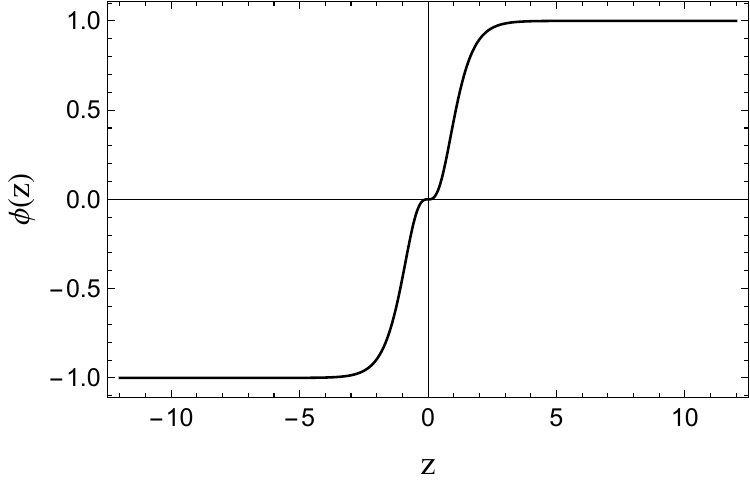}
		\label{fig:k3_phi}
	\end{subfigure}
	\hspace{0.04\textwidth}
	\begin{subfigure}[b]{0.42\textwidth}
		\centering
		\includegraphics[width=\linewidth]{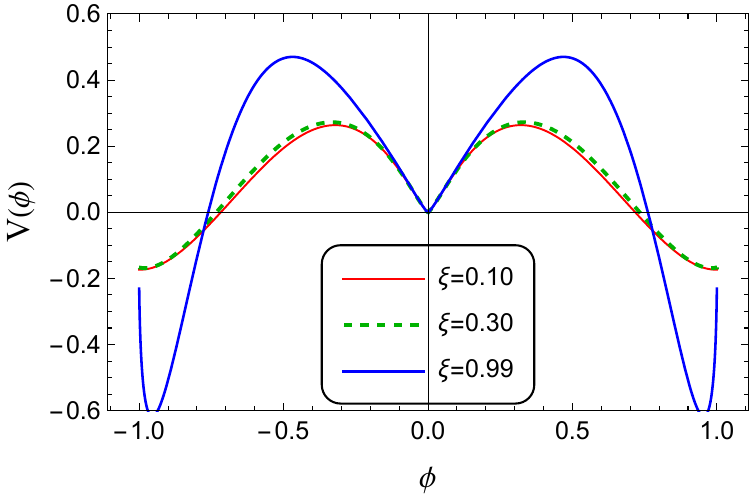}
		\label{fig:k3_Vphi}
	\end{subfigure}
	
	\caption{
		The brane configuration for $k=3$, including the warp factor
		$e^{2A(z)}$, the effective energy density $T_{00}^{\rm eff}(z)$, the scalar field
		$\phi(z)$, and the scalar potential $V(\phi)$. The other parameters
		are set as $\phi_0=1$ and $b=1$.
	}
	\label{fig:k3_background}
\end{figure*}

\begin{figure*}[t]
	\centering
	
	\begin{subfigure}[b]{0.42\textwidth}
		\centering
		\includegraphics[width=\linewidth]{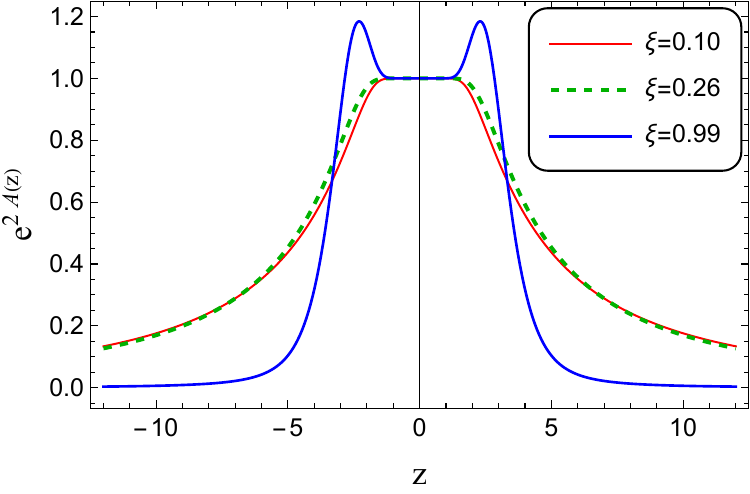}
		\label{fig:k15_warp}
	\end{subfigure}
	\hspace{0.04\textwidth}
	\begin{subfigure}[b]{0.42\textwidth}
		\centering
		\includegraphics[width=\linewidth]{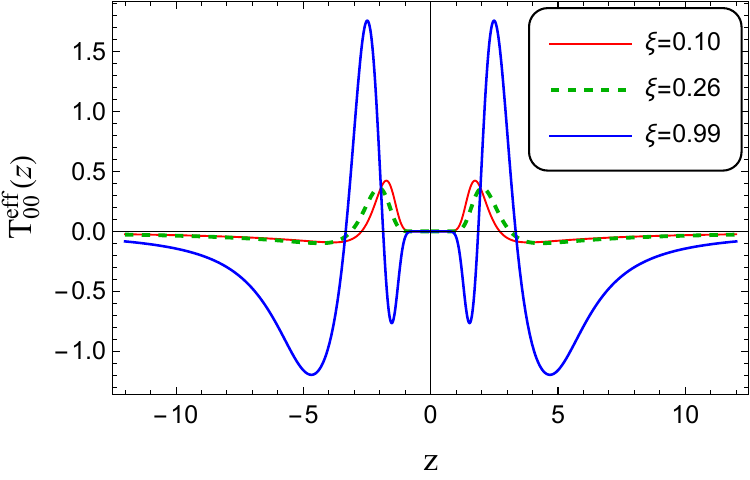}
		\label{fig:k15_T00}
	\end{subfigure}
	
	\vspace{0.2cm}
	
	\begin{subfigure}[b]{0.42\textwidth}
		\centering
		\includegraphics[width=\linewidth]{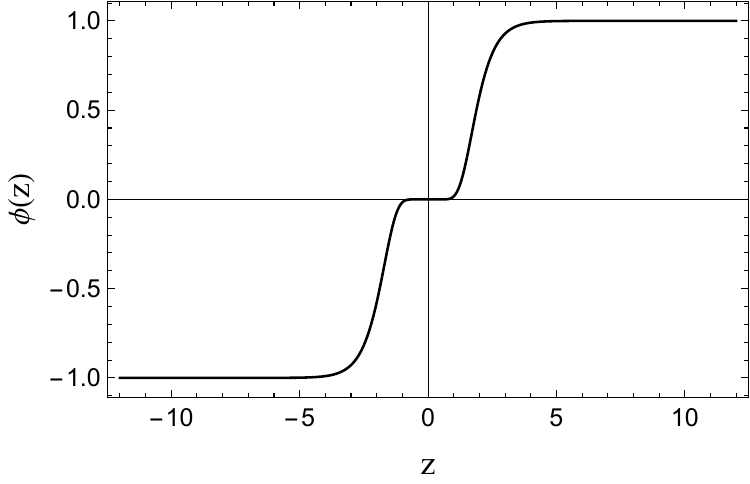}
		\label{fig:k15_phi}
	\end{subfigure}
	\hspace{0.04\textwidth}
	\begin{subfigure}[b]{0.42\textwidth}
		\centering
		\includegraphics[width=\linewidth]{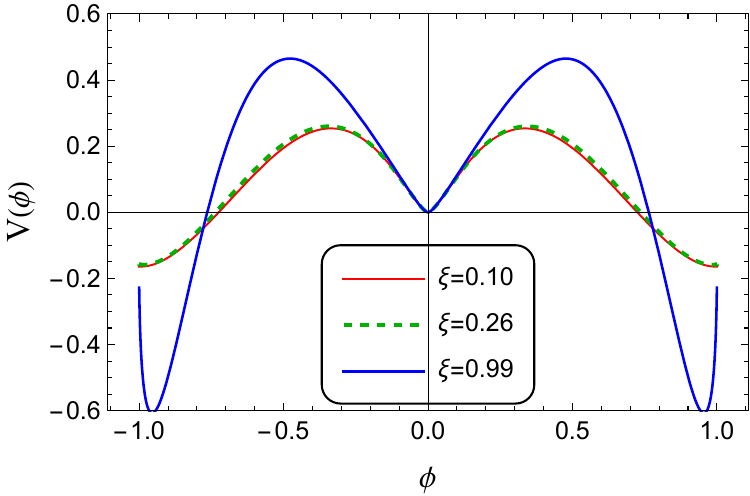}
		\label{fig:k15_Vphi}
	\end{subfigure}
	
	\caption{
		The brane configuration for $k=15$, including the warp factor
		$e^{2A(z)}$, the effective energy density $T_{00}^{\rm eff}(z)$, the scalar field
		$\phi(z)$, and the scalar potential $V(\phi)$. The other parameters
		are set as $\phi_0=1$ and $b=1$.
	}
	\label{fig:k15_background}
\end{figure*}

For $k=15$, the scalar field has the leading behavior
\begin{equation}
	\phi(z)=\phi_0(bz)^{15}
	-5\phi_0(bz)^{17}
	+\mathcal{O}\!\left((bz)^{19}\right).
\end{equation}
The behavior of the warp factor
in the vicinity of $z$ = 0, from Eq.~\eqref{eq:modelB_AZ_equation}, can be expressed as
\begin{equation}
	A(z)=
	\frac{1}{174}\phi_0^2(58\xi-15)(bz)^{30}
	+\mathcal{O}\!\left((bz)^{31}\right).
	\label{eq:modelB_A_k15}
\end{equation}
Therefore, the corresponding critical value is $\xi_0=\frac{15}{58}$.
Compared with the cases \(k=1\) and \(k=3\),
the first correction to \(A(z)\) appears at a much higher order,
indicating that the warp factor varies much more slowly around
\(z=0\).
The effective energy density \(T^{\rm eff}_{00}(z)\) is given by
\begin{equation}
	T^{\rm eff}_{00}(z)
	=
	-3\left[
	A''(z)+A'^2(z)
	\right].
	\label{eq:Teff00_modelB}
\end{equation}
The profiles of the warp factor $e^{2A(z)}$ and the effective energy density 
$T_{00}^{\rm eff}(z)$ for $k=1,3,$ and $15$ are shown in 
Figs.~\ref{fig:k1_background}, \ref{fig:k3_background}, and 
\ref{fig:k15_background}, respectively. 
For $k=1$, increasing $\xi$
narrows the peak of \(e^{2A(z)}\) and splits the effective energy density into two
off-center maxima.  For $k=3$, the double-peak structure is already present at the smallest value of $\xi$ considered and becomes more pronounced as $\xi$ increases. For $k=15$, the two sub-branes are well separated, and their separation increases with $\xi$. These results show that both $\xi$ and $k$ play important roles in determining the internal structure of the brane.

Since the scalar field \eqref{eq:modelB_phi} is monotonic for odd $k$,
the potential can be expressed parametrically in terms of $z$.
Combining Eqs.~\eqref{eq:bg_z_1} and
\eqref{eq:bg_z_2}, and substituting the coupling function
in Eq.~\eqref{eq:F_phi_unified}, the scalar potential can be
written as

\begin{equation}
	V(\phi)
	=
	e^{-2A(z)}
	\left[
	\frac{1}{2}\phi'^2
	-6(1-\xi\phi^2)A'^2
	+8\xi A'\phi\phi'
	\right].
	\label{eq:V_z_modelB}
\end{equation}
As shown in 
Figs.~\ref{fig:k1_background}, \ref{fig:k3_background}, and \ref{fig:k15_background},
  for $k=1$ and small $\xi$, $V(\phi)$ has a $\phi^{4}$-like form with a local maximum at
$\phi=0$; as $\xi$ increases, it flattens and develops a local minimum,
indicating a gradual change toward a \(\phi^6\)-like profile. For $k=3$ and
$k=15$, $V(\phi)$ exhibits a local minimum at $\phi=0$, with
the two symmetric local maxima becoming more pronounced at larger $\xi$.

\section{Gravitational localization and KK resonances}\label{sec:localization}
In this section, we investigate the linear stability of the transverse-traceless tensor perturbations
\cite{Farakos_2007_graviton,Cui_2018}. The perturbed metric can be written as
	\begin{equation}
		ds^2=e^{2A(z)}\left[(\eta_{\mu\nu}+{h}_{\mu\nu})dx^\mu dx^\nu+dz^2\right],
	\end{equation}
	where ${h}_{\mu\nu}$ denotes the tensor perturbation. In the transverse-traceless gauge, the perturbation satisfies
	\begin{equation}
		\partial^\mu {h}_{\mu\nu}=0,\qquad \eta^{\mu\nu}{h}_{\mu\nu}=0.
	\end{equation}
Under these conditions, the tensor fluctuations decouple from the scalar perturbations and satisfy
	\begin{equation}
	\left( \partial_z^2+Q'(z)\partial_z+\Box^{(4)} \right)h_{\mu\nu}=0,
		\label{eq:ttmaster}
	\end{equation}
	where
	\begin{equation}
		Q(z)=3A(z)+\ln F(\phi(z)).
		\label{eq:qdef_22}
	\end{equation}
	The perturbation is further decomposed as
	\begin{equation}
		{h}_{\mu\nu}(x^i, t, z) = e^{-Q(z)/2} \epsilon_{\mu\nu}(x^i) \Phi(t, z),
		\label{eq:timedepansatz}
	\end{equation}
	where $\epsilon_{\mu\nu}(x^i)$ is the three-dimensional tensor eigenfunction satisfying
	\begin{equation}
		\delta^{ij}\partial_i\partial_j\epsilon_{\mu\nu}=-p^2\epsilon_{\mu\nu}.
		\label{eq:emode}
	\end{equation}
	Here \(p\) is the three-dimensional spatial momentum along the brane, and \(\Phi(t,z)\) is the time-dependent mode function of the tensor perturbation.
	Substituting Eq.~(\ref{eq:timedepansatz}) into Eq.~(\ref{eq:ttmaster}), one obtains the time-dependent wave equation
\begin{equation}
	\partial_t^2 \Phi(t,z)-\partial_z^2 \Phi(t,z)
	+V_g(z)\Phi(t,z)=-p^2 \Phi(t,z).
	\label{eq:timewave}
\end{equation}
	Here, the effective potential is
	\begin{equation}
	V_g(z)=\frac{1}{2}Q''(z)+\frac{1}{4}Q'^2(z).
		\label{eq:eff_sch}
	\end{equation}
	Assuming $\Phi(t, z) = e^{-i\omega t} \varphi(z)$,
	Eq.~\eqref{eq:timewave} can be reduced to the Schr\"odinger-like equation
	\begin{equation}
		\left(-\partial_z^2+V_g(z)\right)\varphi(z)=m^2\varphi(z),
		\label{eq:schrodinger}
	\end{equation}
	where$\ m^2=\omega^2-p^2$.
	The Schr\"odinger-like equation (\ref{eq:schrodinger}) can be factorized into a supersymmetric form
	\begin{equation}
		L^\dagger L\,\varphi(z)=m^2\varphi(z),
		\label{eq:susyform}
	\end{equation}
	where the two operators are given by
	\begin{equation}
		L=\partial_z-\frac{1}{2}Q'(z), \qquad
		L^\dagger=-\partial_z-\frac{1}{2}Q'(z).
		\label{eq:operators}
	\end{equation}
	In addition, we can obtain another Schr\"odinger-like equation with the supersymmetric partner potential,
	\begin{equation}
		L L^\dagger \widetilde{\varphi}(z)=\left(-\partial_z^2+V_g^{\mathrm{dual}}(z)\right)\widetilde{\varphi}(z)=m^2\widetilde{\varphi}(z),
		\label{eq:dual_sch}
	\end{equation}
	where
	\begin{equation}
		V_g^{\mathrm{dual}}(z)=-\frac{1}{2}Q''(z)+\frac{1}{4}Q'^2(z).
		\label{eq:dual_potential}
	\end{equation}
	Interestingly, according to supersymmetric quantum mechanics, the Schr\"odinger-like
	equations with the effective potentials \(V_g(z)\) and
	\(V_g^{\rm dual}(z)\) share the same massive KK spectrum and QNMs~\cite{Tan_2025}.
	 Moreover, Eq.~\eqref{eq:susyform} ensures that the tensor KK spectrum contains no tachyonic modes~\cite{Cooper:1994eh}.
	The massless graviton is identified with the zero mode of the Schr\"odinger operator. It satisfies
	\begin{equation}
		L\varphi_0(z)=0,
	\end{equation}
	which gives
	\begin{equation}
		\varphi_0(z)
		=
		Ne^{Q(z)/2}
		=
		Ne^{3A(z)/2}F^{1/2}(\phi(z)),
		\label{eq:zero_mode}
	\end{equation}
	where \(N\) is the normalization constant.
	The localization of the massless graviton requires the zero mode to
	be normalizable,
	\begin{equation}
		\int_{-\infty}^{+\infty}
		|\varphi_0(z)|^2 dz
		=
		N^2
		\int_{-\infty}^{+\infty}
		e^{3A(z)}F(\phi(z))\,dz
		=1 .
		\label{eq:zero_mode_normalization}
	\end{equation}
	Therefore, the normalization condition is equivalent to
	\begin{equation}
		\int_{-\infty}^{+\infty}
		e^{3A(z)}F(\phi(z))\,dz <\infty .
		\label{eq:zero_mode_normalization_1}
	\end{equation}
We now apply this formalism to the two models and analyze the corresponding localization properties and possible resonant KK modes.
	
		\subsection{Model A}
		\label{subsec:modelA_localization}
	
\begin{figure*}[htbp]
	\centering
	\includegraphics[width=0.42\textwidth]{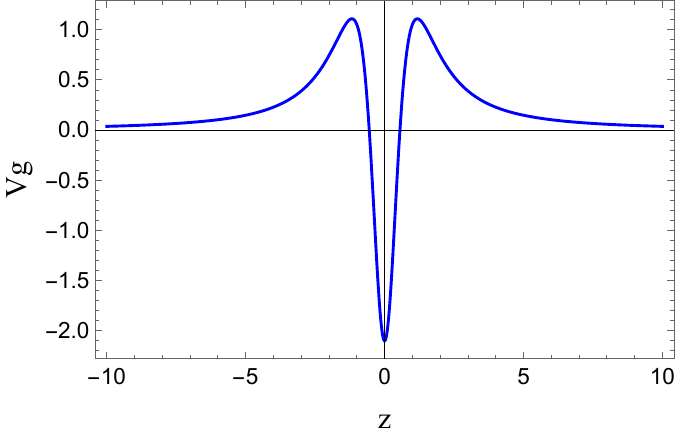}
     \hspace{0.06\textwidth}
	\includegraphics[width=0.42\textwidth]{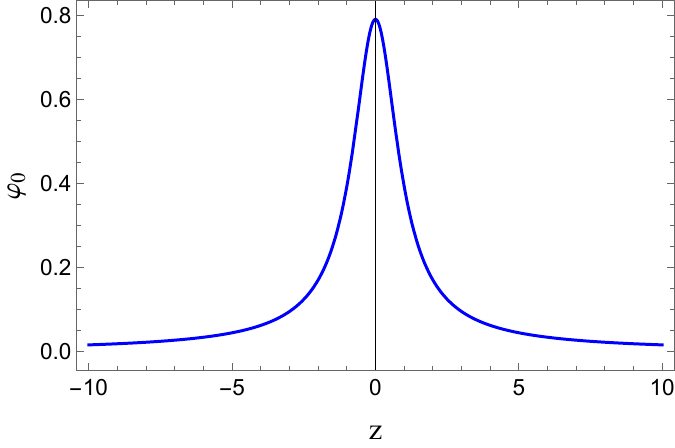}
	\caption{
		(a) Tensor effective potential \(V_g(z)\);
		(b) normalized graviton zero mode \(\varphi_0(z)\)
		for Model A with \(\xi=1/7\) and \(\kappa=1\).
	}
	\label{fig:modelA_Vg_zero}
\end{figure*}

We begin with Model~A, reviewed in Sec.~\ref{subsec:modelA}. 
	In the present work, we focus on the special case $\xi=1/7$, for which the tensor-perturbation effective potential admits an analytical expression, as given in Ref.~\cite{Liu_2012}. Here the background solution is written in the coordinate $y$, while the effective potential is expressed in the conformal coordinate $z$. In this case, the coordinate transformation can be performed explicitly, and hence the effective potential can be written analytically as a function of $z$.
	The corresponding effective potential  for tensor perturbations is
	\begin{equation}
		V_{g}(z)=\frac{15\kappa^{2}\left(-14+37\kappa^{2}z^{2}+28\kappa^{4}z^{4}+4\kappa^{6}z^{6}\right)}
		{4\left(5+7\kappa^{2}z^{2}+2\kappa^{4}z^{4}\right)^{2}}.
		\label{eq:vgz}
	\end{equation}
	As shown in Fig.~\ref{fig:modelA_Vg_zero}(a), the effective potential
	exhibits a volcano-like profile and approaches zero as
	$|z|\to\infty$.
	
	The normalized zero mode is given by
	\begin{equation}
		\varphi_{0}(z)=\sqrt{\frac{\kappa}{8}}\,\frac{\sqrt{5+2\kappa^{2}z^{2}}}{\left(1+\kappa^{2}z^{2}\right)^{5/4}},
		\label{eq:modelA_zero_mode}
	\end{equation}
whose profile is shown in Fig.~\ref{fig:modelA_Vg_zero}(b).
It satisfies the normalization condition
\begin{equation}
	\int_{-\infty}^{+\infty}
	\left|\varphi_{0}(z)\right|^{2}dz
	=
	\frac{\kappa}{8}
	\int_{-\infty}^{+\infty}
	\frac{5+2\kappa^{2}z^{2}}
	{\left(1+\kappa^{2}z^{2}\right)^{5/2}}\,dz
	=1.
	\label{eq:modelA_zero_normalization}
\end{equation}
Therefore, the graviton zero mode is normalizable, and the massless
graviton is localized on the brane. No pronounced resonance peaks are found for Model A at $\xi=1/7$.

	\subsection{Model B}
	\label{subsec:modelB_localization}

\begin{figure}[htbp]
	\centering
	
	\begin{subfigure}[b]{0.32\textwidth}
		\centering
		\includegraphics[width=\linewidth]{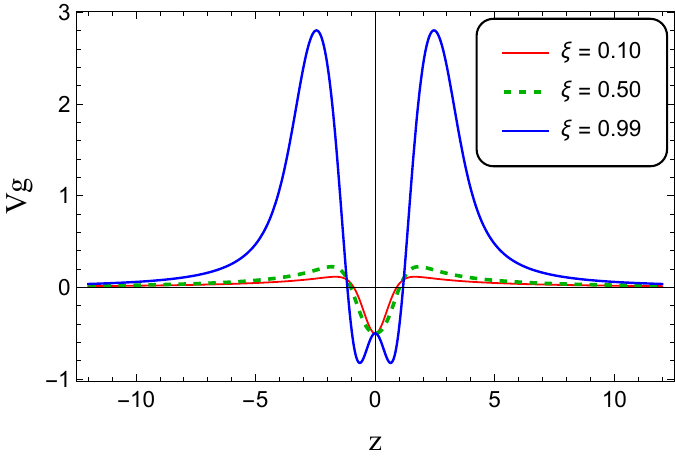}
		\caption{\(k=1\).}
		\label{fig:modelB_vgz_k1}
	\end{subfigure}
	\hfill
	\begin{subfigure}[b]{0.32\textwidth}
		\centering
		\includegraphics[width=\linewidth]{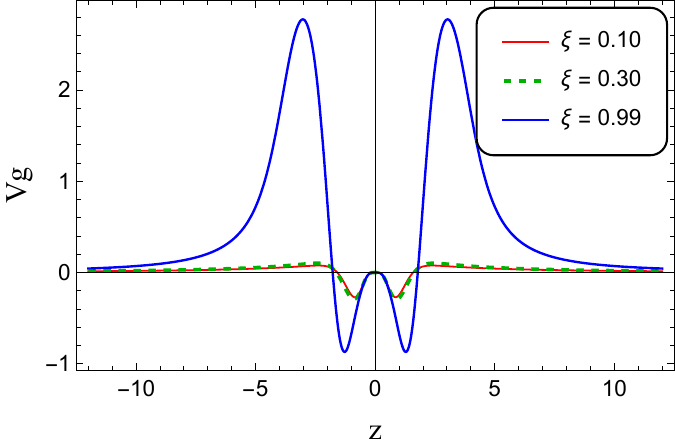}
		\caption{\(k=3\).}
		\label{fig:modelB_vgz_k3}
	\end{subfigure}
	\hfill
	\begin{subfigure}[b]{0.32\textwidth}
		\centering
		\includegraphics[width=\linewidth]{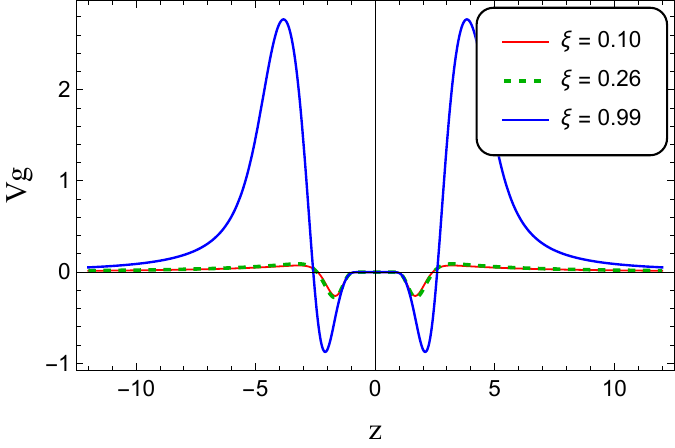}
		\caption{\(k=15\).}
		\label{fig:modelB_vgz_k15}
	\end{subfigure}
	
\caption{
	The effective potential \(V_g(z)\) in Model B for
	(a) \(k=1\), (b) \(k=3\), and (c) \(k=15\), with several
	values of the non-minimal coupling parameter \(\xi\).
	The other parameters are fixed as \(\phi_0=1\) and \(b=1\).
}
	\label{fig:modelB_vgz_k}
\end{figure}

\begin{figure}[htbp]
	\centering
	
	\begin{subfigure}[b]{0.32\textwidth}
		\centering
		\includegraphics[width=\linewidth]{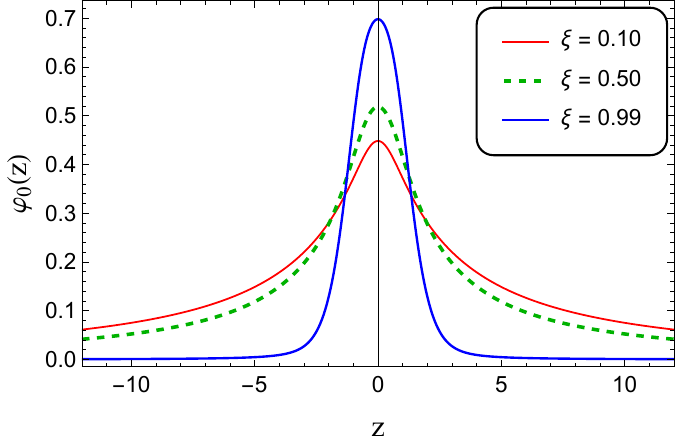}
		\caption{\(k=1\).}
		\label{fig:modelB_zero_k1}
	\end{subfigure}
	\hfill
	\begin{subfigure}[b]{0.32\textwidth}
		\centering
		\includegraphics[width=\linewidth]{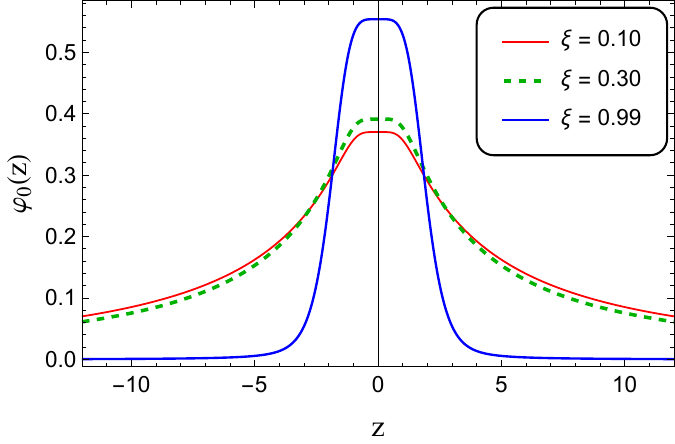}
		\caption{\(k=3\).}
		\label{fig:modelB_zero_k3}
	\end{subfigure}
	\hfill
	\begin{subfigure}[b]{0.32\textwidth}
		\centering
		\includegraphics[width=\linewidth]{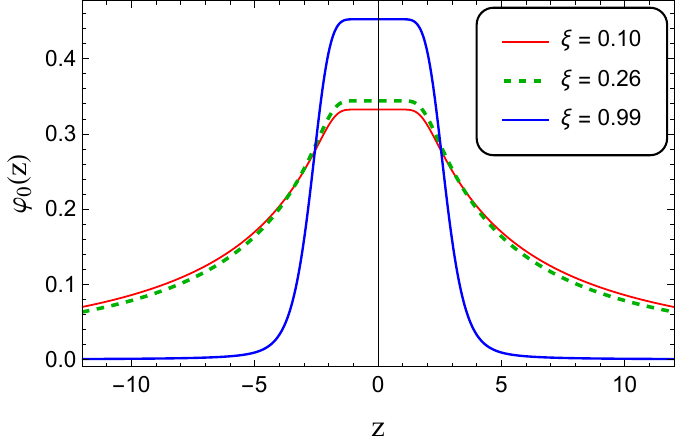}
		\caption{\(k=15\).}
		\label{fig:modelB_zero_k15}
	\end{subfigure}
	
	\caption{
		The normalized graviton zero mode \(\varphi_0(z)\) in Model B
		for (a) \(k=1\), (b) \(k=3\), and (c) \(k=15\), with several
		values of the non-minimal coupling parameter \(\xi\).
		The other parameters are fixed as \(\phi_0=1\) and \(b=1\).
	}
	\label{fig:modelB_zero_k}
\end{figure}

For Model B, the effective potential of the gravitational KK modes
and the corresponding graviton zero mode are obtained from
Eqs.~\eqref{eq:eff_sch} and \eqref{eq:zero_mode}, respectively.
The results for \(k = 1, 3, 15\) are shown in
Figs.~\ref{fig:modelB_vgz_k} and \ref{fig:modelB_zero_k}.

By considering the asymptotic behavior of the warp factor and the
background scalar field, the effective potential satisfies
\begin{equation}
	V_g(z\to\pm\infty)\to0,
	\label{eq:modelB_Vg_asymptotic}
\end{equation}
which corresponds to a volcano-like potential. For fixed $k$,
increasing $\xi$ deepens the central potential well and raises the
two symmetric barriers, producing a pronounced double-barrier
structure at $\xi=0.99$. For fixed $\xi$, the separation between the two potential barriers grows with $k$, consistent with the increasing distance between the two sub-branes.
The asymptotic behavior of the graviton zero mode is
\begin{equation}
	\varphi_0(z\to\pm\infty)
	\propto
	\sqrt{1-\xi\phi_0^2}
	\left(\frac{1}{b|z|}\right)^{3/2}.
	\label{eq:modelB_zero_asymptotic}
\end{equation}
Thus, the normalization condition~\eqref{eq:zero_mode_normalization_1}
is satisfied since
\begin{equation}
	\int_{1}^{+\infty}
	|\varphi_0(z)|^2\,dz
	\propto
	\frac{1-\xi\phi_0^2}{b^3}
	\int_{1}^{+\infty}\frac{dz}{z^3}
	=
	\frac{1-\xi\phi_0^2}{2b^3}
	<\infty.
	\label{eq:modelB_zero_normalizability}
\end{equation}
Therefore, the graviton zero mode is normalizable and localized on the brane for all the parameter values considered. For $k=1$, the zero mode is mainly localized around the brane center for $\xi<\xi_0$, whereas for $\xi$ close to or larger
than $\xi_0$, it is localized in the region between the two sub-branes. For $k=3$, the zero mode is localized mainly in the region between the two sub-branes. A similar behavior is found
for $k=15$, while the zero mode profile extends over a wider region between the two sub-branes and becomes nearly flat-topped.

	\begin{figure}[htbp]
		\centering
		
		\includegraphics[width=0.32\textwidth]{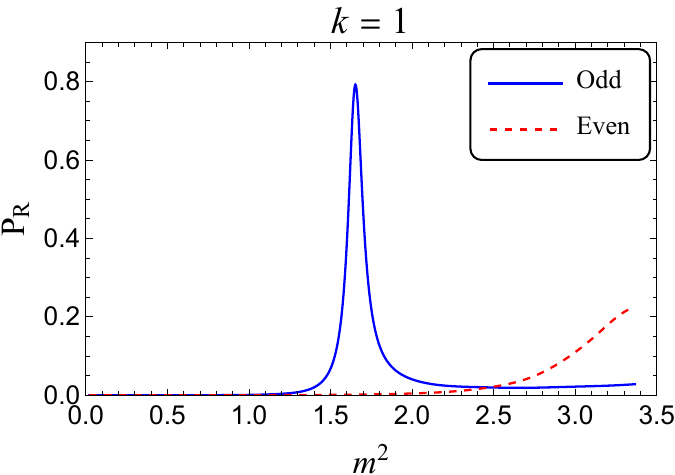}
		\hfill
		\includegraphics[width=0.32\textwidth]{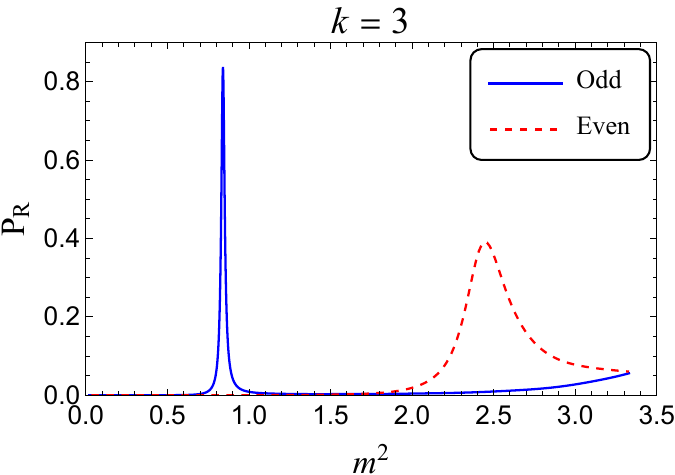}
		\hfill
		\includegraphics[width=0.32\textwidth]{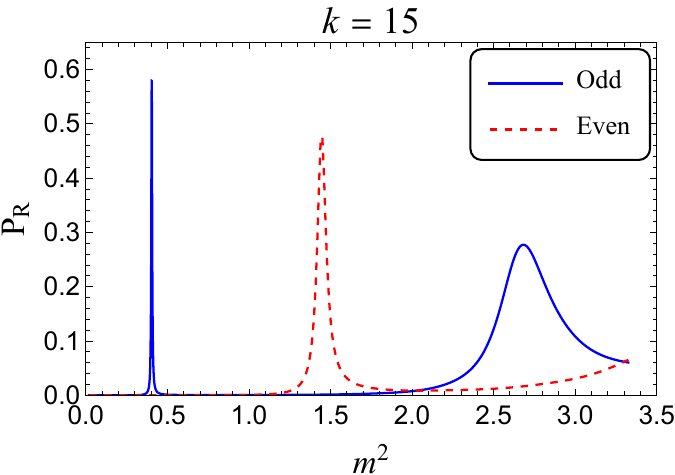}
		
	\caption{
		The relative probabilities \(P_R(m^2)\) of the odd- and
		even-parity massive KK modes for
		(a) \(k=1\), (b) \(k=3\), and (c) \(k=15\).
		The parameters are fixed as
		\(\xi=0.99\), \(\phi_0=1\), and \(b=1\).
		}
		\label{fig:relative_probability_k}
	\end{figure}
A volcano-like effective potential may support quasi-localized resonant states.
To identify the resonant KK modes, we use the relative probability
method. For each mass \(m\), the wave function \(\varphi(z)\) is obtained by solving the 
Schrödinger-like equation \eqref{eq:schrodinger}. The relative 
probability is defined as
\begin{equation}
	P_R(m^2)
	=
	\frac{
		\displaystyle \int_{-z_b}^{z_b} |\varphi(z)|^2 dz
	}{
		\displaystyle \int_{-z_{\max}}^{z_{\max}} |\varphi(z)|^2 dz
	},
	\label{eq:relative_probability}
\end{equation}
where \(2z_b\) is approximately the width of the thick brane, and
\(z_{\max}=10z_b\). For KK modes with \(m^2\gg V_{\max}\), the wave functions
asymptotically behave as plane waves, and the corresponding relative
probability \(P_R(m^2)\) approaches \(1/10\), where \(V_{\max}\)
denotes the maximum of the corresponding effective potential. Each peak of \(P_R(m^2)\) identifies a resonant
KK mode. Following Refs.~\cite{Liu_2009_resonance,Liu_2009_fermion},
the lifetime of a resonance is estimated as
\(\tau\sim\Gamma^{-1}\), where \(\Gamma=\delta m\) is the full width at half maximum of the resonant peak.
For smaller $\xi$ (e.g., $\xi=0.10$ and $0.30$), only a
few resonant KK modes are identified, or none at all.  We
therefore focus on $\xi=0.99$, where the double-barrier structure
is most pronounced, in the following resonance analysis.
  As shown in Fig.~\ref{fig:relative_probability_k}, the number of resonant KK modes increases
 with $k$. Their squared masses, masses, widths, and estimated lifetimes are listed
 in Table~\ref{tab:modelB_resonances}.

\begin{table}[htbp]
	\centering
	\caption{Mass spectrum, full widths at half maximum  $\Gamma$, and lifetime $\tau$ of the gravitational KK resonances for Model B with $\xi=0.99$, $\phi_0=1$, and $b=1$.}
	\label{tab:modelB_resonances}
	\begin{tabular}{cccccc}
		\toprule
		$k$ & $n$ & $m_n^2$ & $m_n$ & $\Gamma$ & $\tau$ \\
		\midrule
		$1$ & $1$ & $1.634424$ & $1.27845$ & $0.041632$ & $24.02$ \\
		\midrule
		\multirow{2}{*}{$3$}
		& $1$ & $0.831835$ & $0.91205$ & $0.013373$ & $74.78$ \\
		& $2$ & $2.433646$ & $1.56001$ & $0.104536$ & $9.57$ \\
		\midrule
		\multirow{3}{*}{$15$}
		& $1$ & $0.403800$ & $0.63545$ & $0.004820$ & $207.47$ \\
		& $2$ & $1.449700$ & $1.20403$ & $0.027663$ & $36.15$ \\
		& $3$ & $2.681800$ & $1.63762$ & $0.122853$ & $8.14$ \\
		\bottomrule
	\end{tabular}
\end{table}

\begin{figure}[htbp]
	\centering
	
	\begin{subfigure}[b]{0.31\textwidth}
		\centering
		\includegraphics[width=\textwidth]{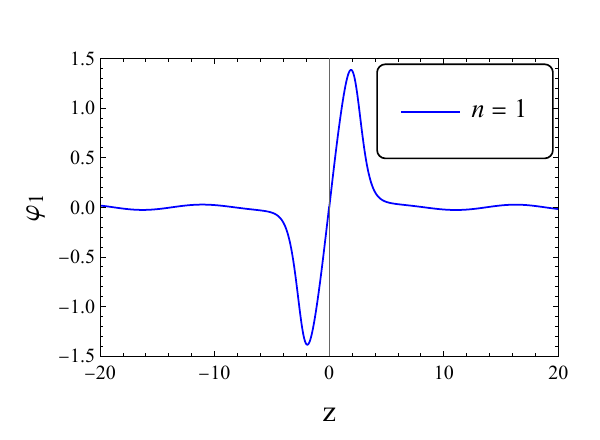}
		\label{fig:wave_k15_n1}
	\end{subfigure}
	\hfill
	\begin{subfigure}[b]{0.31\textwidth}
		\centering
		\includegraphics[width=\textwidth]{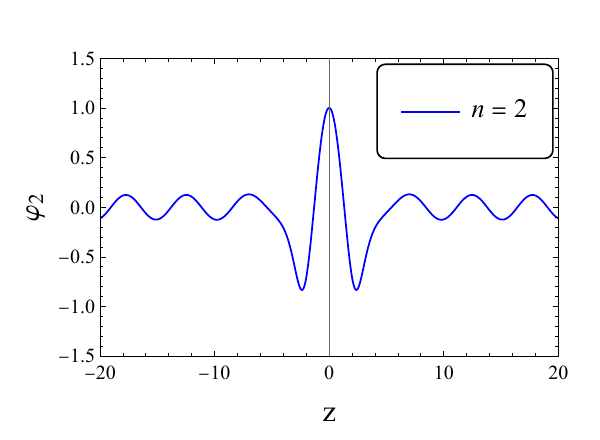}
		\label{fig:wave_k15_n2}
	\end{subfigure}
	\hfill
	\begin{subfigure}[b]{0.31\textwidth}
		\centering
		\includegraphics[width=\textwidth]{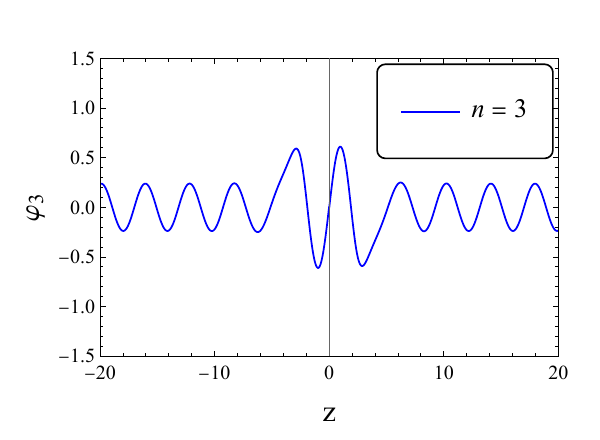}
		\label{fig:wave_k15_n3}
	\end{subfigure}
		\caption{
			The wave functions of the three resonant KK modes in
			Model B for \(k=15\): (a) \(n=1\), (b) \(n=2\), and
			(c) \(n=3\). The parameters are fixed as
			\(\xi=0.99\), \(\phi_0=1\), and \(b=1\).
		}
	\label{fig:wavefunction_k15}
\end{figure}
\begin{figure}[htbp]
	\centering
	
	\begin{subfigure}[b]{0.42\textwidth}
		\centering
		\includegraphics[width=\linewidth]{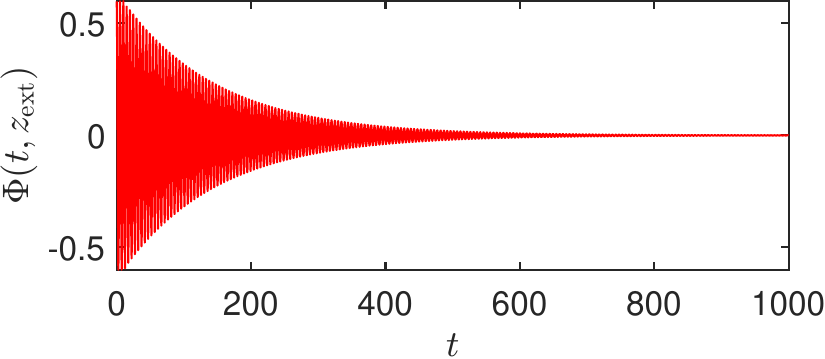}
		\caption{\(k=1\)}
		\label{fig:phi_k_1}
	\end{subfigure}
\hspace{0.06\textwidth}
	\begin{subfigure}[b]{0.42\textwidth}
		\centering
		\includegraphics[width=\linewidth]{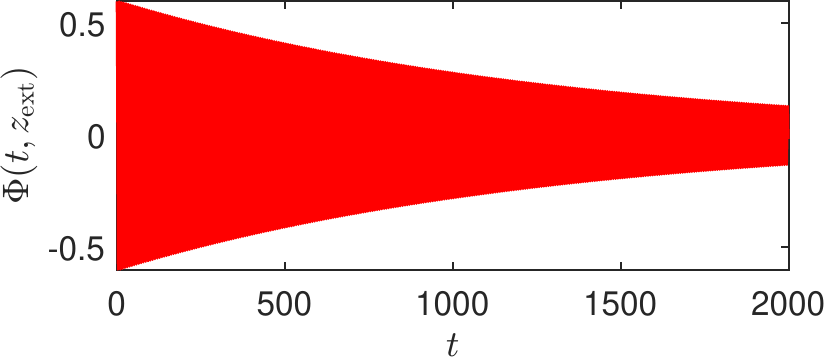}
		\caption{\(k=3\)}
		\label{fig:phi_k_3}
	\end{subfigure}
	
	\vspace{0.35cm}
	
	\begin{subfigure}[b]{0.42\textwidth}
		\centering
		\includegraphics[width=\linewidth]{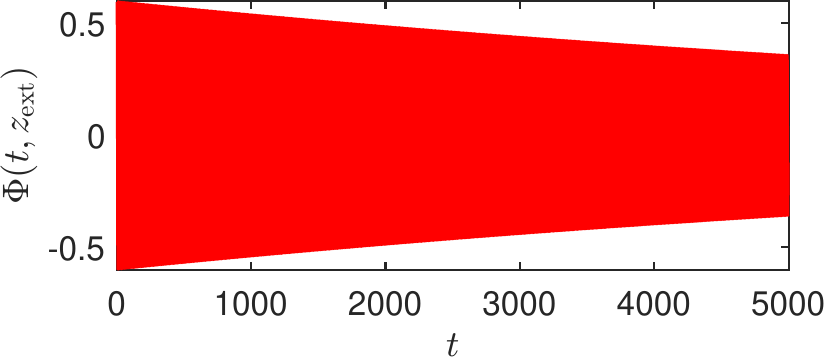}
		\caption{\(k=15\)}
		\label{fig:phi_k_15}
	\end{subfigure}
	
	\caption{
		Time evolution of the first resonant KK mode in
		Model B for \(k=1\), \(3\), and \(15\), recorded at the fixed
		extraction point \(z_{\rm ext}=1\).
	\label{fig:time_evolution_k}
}
\end{figure}

Using the wave functions of the resonant KK modes obtained above
as the initial data, we further investigate the time evolution of
these resonant states, following the numerical method of
Ref.~\cite{Tan_2023_scalarRes}. We work in the rest frame along the
brane by setting the three-dimensional momentum to \(p=0\), in which
case Eq.~\eqref{eq:timewave} reduces to
\begin{equation}
	\left[\partial_t^2-\partial_z^2+V_g(z)\right]\Phi(t,z)=0.
	\label{eq:time_evolution_modelB}
\end{equation}
Since massive KK modes can propagate away from the brane into the asymptotic region, 
maximally dissipative boundary conditions are imposed at the numerical boundaries
\cite{Megevand_2007},
\begin{equation}
	\begin{aligned}
		\left.(\partial_t+\partial_z)\Phi(t,z)\right|_{z=z_{\max}} &= 0,\\
		\left.(\partial_t-\partial_z)\Phi(t,z)\right|_{z=-z_{\max}} &= 0.
	\end{aligned}
	\label{eq:max_dissipative_bc}
\end{equation}
Here, \(z_{\max}\) is chosen sufficiently large that
\(V_g(z)\) is negligible near the boundaries. The $n$th resonant KK wave function $\varphi_n(z)$ is
taken as the initial configuration
\begin{equation}
	\Phi(0,z)=\varphi_n(z),
	\qquad
	\partial_t\Phi(0,z)=0.
	\label{eq:resonance_initial_data}
\end{equation}
For $k=15$, the spatial profiles of the three resonant KK modes identified from the relative-probability spectrum are displayed in Fig.~\ref{fig:wavefunction_k15}. 
Equation~\eqref{eq:time_evolution_modelB} is solved numerically using
fourth-order finite differences in space and the fourth-order Runge--Kutta
method in time. We extract the time series of the resonance amplitude
$\Phi(t,z_{\rm ext})$ at a fixed position $z_{\rm ext}=1$. 
Figure~\ref{fig:time_evolution_k} shows the time evolution of the first resonant KK mode for $k=1$, $3$, and $15$.
The oscillation amplitude gradually decreases with time, while the decay becomes
slower as $k$ increases, indicating that the first resonant KK mode remains concentrated near the brane for a longer time.
By fitting the exponential envelopes of
\(\Phi(t,z_{\rm ext})\), we obtain the amplitude decay rates
\(\gamma^{\rm res}\) of the resonant KK modes. The fitted results
are summarized in Table~\ref{tab:modelB_time_decay}. 

\begin{figure}[htbp]
	\centering
	
	\begin{subfigure}[b]{0.31\textwidth}
		\centering
		\includegraphics[width=\textwidth]{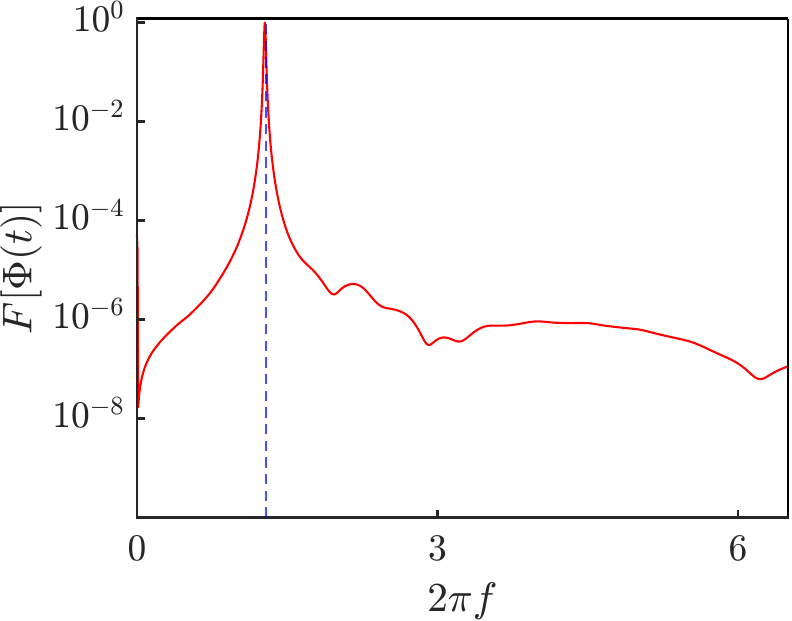}
		\caption{\(k=1\)}
		\label{fig:fourier_k1}
	\end{subfigure}
	\hfill
	\begin{subfigure}[b]{0.31\textwidth}
		\centering
		\includegraphics[width=\textwidth]{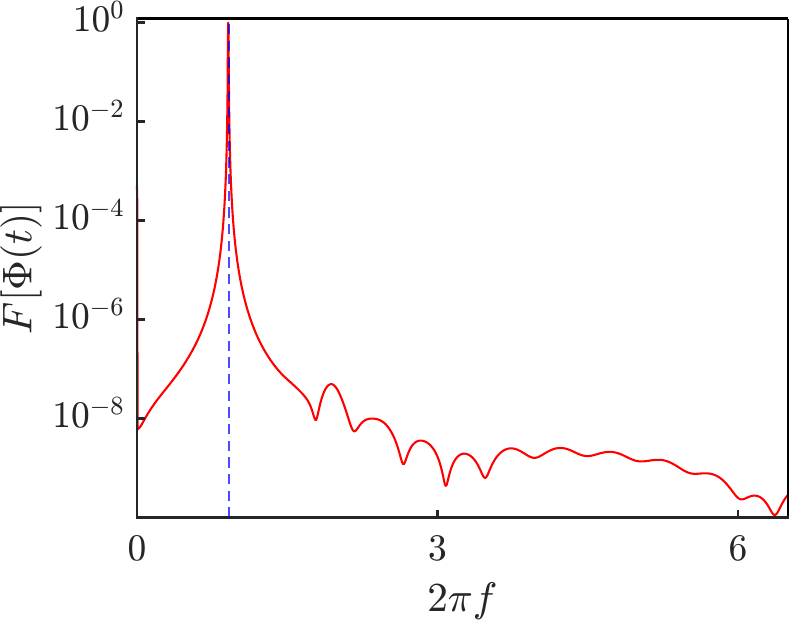}
		\caption{\(k=3\)}
		\label{fig:fourier_k3}
	\end{subfigure}
	\hfill
	\begin{subfigure}[b]{0.31\textwidth}
		\centering
		\includegraphics[width=\textwidth]{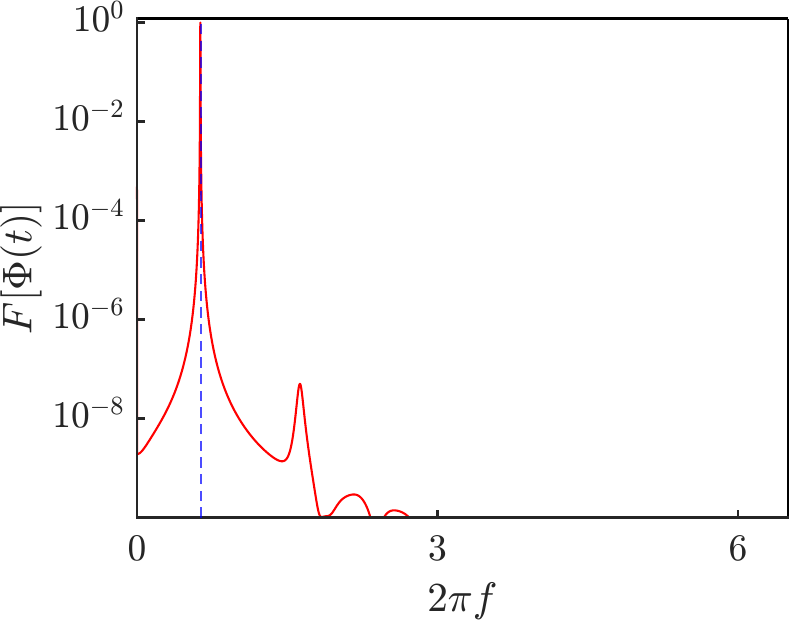}
		\caption{\(k=15\)}
		\label{fig:fourier_k15}
	\end{subfigure}
	
\caption{
	Fourier spectra obtained from the time evolution signals.
	The blue dotted lines indicate the masses of the first resonant KK modes.
}
	\label{fig:fourier_first_resonance_k}
\end{figure}

\begin{table}[htbp]
	\centering
	\caption{The decay rates
		\(\gamma^{\rm res}\) of the resonant KK modes for
		different values of \(k\). The symbol ``--'' indicates
		that no corresponding resonant KK mode is present.}
	\label{tab:modelB_time_decay}
\begin{tabular}{cccc}
	\toprule
	\(n\)
	& \(\gamma^{\rm res}(k=1)\)
	& \(\gamma^{\rm res}(k=3)\)
	& \(\gamma^{\rm res}(k=15)\) \\
	\midrule
	1
	& \(7.36\times10^{-3}\)
	& \(7.59\times10^{-4}\)
	& \(9.33\times10^{-5}\) \\
	2
	& --
	& \(2.801\times10^{-2}\)
	& \(2.887\times10^{-3}\) \\
	3
	& --
	& --
	& \(3.415\times10^{-2}\) \\
	\bottomrule
\end{tabular}
\end{table}
To extract the characteristic oscillation frequencies, we perform a discrete Fourier
transform of the time evolution data recorded at $z_{\rm ext}=1$. The Fourier amplitude is
defined as
\begin{equation}
	F[\Phi(t)](f)
	=
	A
	\left|
	\sum_{j}
	\Phi(t_j,z_{\rm ext})
	\exp\left(-2\pi i f t_j\right)
	\right|,
	\label{eq:discrete_fourier}
\end{equation}
where \(t_j\) denotes the discrete sampling time, \(f\) is the
ordinary frequency, and \(A\) is a normalization constant. Before performing the discrete Fourier transform, the mean value is subtracted from the time series and a Hann window is applied to reduce spectral leakage.
The Fourier spectra are shown in
Fig.~\ref{fig:fourier_first_resonance_k}. For each value of \(k\), the
mass of the first resonant KK mode obtained from the
relative probability spectrum agrees well with the corresponding peak
of the Fourier spectrum. Since the evolution is performed with
\(p=0\), one has \(\omega=m\), where \(\omega=2\pi f\). Thus, the
angular frequency \(\omega_{\rm peak}=2\pi f_{\rm peak}\) associated
with the Fourier peak directly gives the mass of the corresponding
resonant KK mode. As \(k\) increases, the peak shifts toward a lower
frequency, consistently with the decrease in the mass of the first
resonant mode listed in Table~\ref{tab:modelB_resonances}.

\section{Quasinormal modes of thick branes}\label{sec:qnm}
In this section, we investigate the QNMs of the two
non-minimally coupled thick brane models. The QNMs are obtained by solving the Schr\"odinger-like equation
for tensor perturbations with purely outgoing boundary conditions~\cite{Konoplya_2011}. For
each model, the results are validated using independent
frequency-domain and time-domain methods.

\subsection{Quasinormal modes of Model A}
\label{subsec:qnm_modelA}

For Model A, we calculate the QNFs for the \(\xi=1/7\) case considered
in Sec.~\ref{subsec:modelA_localization}. Substituting the corresponding analytical
expression for \(Q(z)\) into the supersymmetric partner
equation~\eqref{eq:dual_potential} gives the dual effective potential

\begin{equation}
	V_g^{\text{dual}}(z) = \frac{3\kappa^{2} \left( 70 + 109\kappa^{2}z^{2} + 28 \kappa^{4}z^{4} + 4\kappa^{6}z^{6} \right)}{4\left(5+7\kappa^{2}z^{2}+2\kappa^{4}z^{4}\right)^{2}}.
	\label{eq:vgzdual_k}
\end{equation}
The corresponding equation in Schr\"odinger form is
\begin{equation}
	-\partial_z^2\widetilde{\varphi}(z)
	+\left(
	\frac{3\kappa^{2} \left( 70 + 109\kappa^{2}z^{2} + 28 \kappa^{4}z^{4} + 4\kappa^{6}z^{6} \right)}{4\left(5+7\kappa^{2}z^{2}+2\kappa^{4}z^{4}\right)^{2}}
	- m^2
	\right)
	\widetilde{\varphi}(z)=0.
	\label{eq:sch_explicit}
\end{equation}
\begin{figure*}[htbp]
	\centering
	\includegraphics[width=0.42\textwidth]{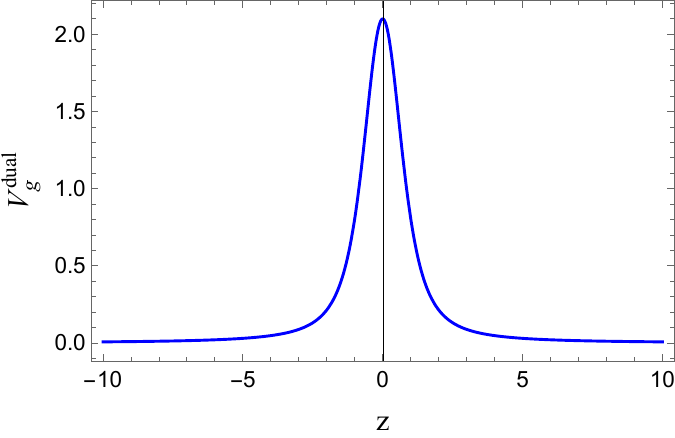}
	\caption{The dual effective potential \(V_g^{\mathrm{dual}}(z)\) given by Eq.~(\ref{eq:vgzdual_k}).}
	\label{fig:dual}
\end{figure*}

As shown in Fig.~\ref{fig:dual}, the dual potential has a single potential barrier.
The boundary conditions for the Schrödinger-like equation can be set as
\begin{equation}
\widetilde{\varphi}(z)\sim
	\begin{cases}
		e^{imz}, & z\rightarrow +\infty,\\[4pt]
		e^{-imz}, & z\rightarrow -\infty.
	\end{cases}
	\label{eq:modelA_bc}
\end{equation}
\subsubsection{Calculation of QNMs in the frequency domain}

First, we employ the asymptotic iteration method (AIM) to compute the QNFs of Model A. The AIM was originally introduced for solving second-order homogeneous linear differential equations~\cite{Ciftci_2003} and has subsequently been applied to QNM calculations~\cite{Cho_2010,Cho_2012_AIM_review}. For completeness, we briefly review its basic idea. Consider the following second-order homogeneous linear
differential equation satisfied by $y(x)$,
\begin{equation}
	y''(x)=\lambda_0(x)y'(x)+s_0(x)y(x),
	\label{eq:AIM1}
\end{equation}
where $\lambda_0(x)\neq 0$ and $s_0(x)$ are sufficiently smooth functions. The AIM is based on the recursive structure generated by successive
differentiations of this equation. By differentiating Eq.~\eqref{eq:AIM1} with respect to $x$, one obtains
\begin{equation}
	y'''(x)=\lambda_1(x)y'(x)+s_1(x)y(x),
	\label{eq:AIM2}
\end{equation}
where
\begin{align}
	\lambda_1(x)
	&=
	\lambda_0'(x)+s_0(x)+\lambda_0^2(x),
	\label{eq:aim_lambda1}\\
	s_1(x)
	&=
	s_0'(x)+s_0(x)\lambda_0(x).
	\label{eq:aim_s1}
\end{align}
Proceeding iteratively, the derivatives of orders \(n+1\) and \(n+2\) of $y(x)$ can be written as
\begin{align}
	y^{(n+1)}(x)
	&=
	\lambda_{n-1}(x)y'(x)
	+s_{n-1}(x)y(x),
	\label{eq:aim_nplus1}\\
	y^{(n+2)}(x)
	&=
	\lambda_n(x)y'(x)
	+s_n(x)y(x),
	\label{eq:aim_nplus2}
\end{align}
where the functions $\lambda_n(x)$ and $s_n(x)$ satisfy the recursion relations
\begin{align}
	\lambda_n(x)
	&= \lambda'_{n-1}(x)+s_{n-1}(x)
	+\lambda_0(x)\lambda_{n-1}(x),
	\label{eq:AIM8}\\
	s_n(x)
	&= s'_{n-1}(x)+s_0(x)\lambda_{n-1}(x).
	\label{eq:AIM9}
\end{align}
For sufficiently large $n$, the asymptotic iteration condition implies that
\begin{equation}
	\frac{s_n(x)}{\lambda_n(x)}=\frac{s_{n-1}(x)}{\lambda_{n-1}(x)}\equiv \beta(x),
	\label{eq:AIM10}
\end{equation}
which can equivalently be written as
\begin{equation}
	s_n(x)\lambda_{n-1}(x)-s_{n-1}(x)\lambda_n(x)=0.
	\label{eq:AIM11}
\end{equation}
The QNFs can be obtained from this quantization condition. However, the direct evaluation of the recursion relations becomes increasingly cumbersome for large iteration number $n$. To improve the computational efficiency and numerical accuracy, Cho \textit{et al.}~\cite{Cho_2010}  proposed an improved AIM based on the Taylor expansion method. In this approach, $\lambda_n(x)$ and $s_n(x)$ are expanded around a chosen point $\chi$ as
\begin{align}
	\lambda_n(x)
	&=
	\sum_{i=0}^{\infty}
	c_n^i(x-\chi)^i,
	\label{eq:aim_lambda_taylor}\\
	s_n(x)
	&=
	\sum_{i=0}^{\infty}
	d_n^i(x-\chi)^i,
	\label{eq:AIM13}
\end{align}
where $c_n^i$ and $d_n^i$ denote the $i$-th Taylor coefficients of $\lambda_n(x)$ and $s_n(x)$, respectively. Substituting these expansions into the recursion relations yields
\begin{align}
	c_n^i
	&=
	(i+1)c_{n-1}^{i+1}
	+d_{n-1}^i
	+\sum_{k=0}^{i}c_0^k c_{n-1}^{i-k},
	\label{eq:aim_c_recursion}\\
	d_n^i
	&=
	(i+1)d_{n-1}^{i+1}
	+\sum_{k=0}^{i}d_0^k c_{n-1}^{i-k}.
	\label{eq:aim_d_recursion}
\end{align}
Now, the quantization condition can be rewritten as
\begin{equation}
	d_n^0 c_{n-1}^0 - d_{n-1}^0 c_n^0 = 0.
	\label{eq:AIM_quantization_final}
\end{equation}
In this way, the original quantization condition is reduced to a set of recursive algebraic relations without the explicit evaluation of higher-order derivatives, which significantly improves the efficiency of the numerical computation.

Since Eq.~\eqref{eq:sch_explicit} contains no term involving the first
derivative, we introduce the coordinate transformation
$u=\frac{\sqrt{4\kappa^{2}z^{2}+1}-1}{2\kappa z}$ to cast the equation into a form suitable for the AIM. The Schr\"odinger-like equation~\eqref{eq:sch_explicit} then becomes
\begin{equation}
	\begin{aligned}
		&\frac{(u^2-1)^3}{(1+u^2)^3}
		\Bigl[
		2u(3+u^2)\widetilde{\varphi}'(u)
		+(u^4-1)\widetilde{\varphi}''(u)
		\Bigr]
		\\[4pt]
		&\quad
		+\Biggl[
		\frac{m^2}{\kappa^2}
		-\frac{3(u^{2}-1)^{2}}
		{4(1-u^{2}+u^{4})^{2}(5-8u^{2}+5u^{4})^{2}}
		\Bigl(
		70-311u^{2}+642u^{4}-798u^{6}
		\\
		&\hspace{6.2cm}
		+642u^{8}-311u^{10}+70u^{12}
		\Bigr)
		\Biggr]\widetilde{\varphi}(u)=0 .
	\end{aligned}
	\label{eq:sch_u_raw}
\end{equation}
Here, the new coordinate variable is restricted to the finite interval $-1<u<1$. Accordingly, the boundary conditions in Eq.~\eqref{eq:modelA_bc} can be rewritten in terms of $u$ as
\begin{equation}
	\widetilde{\varphi}(u)\propto
	\begin{cases}
		e^{-\frac{i m/\kappa}{2u-2}}, & u\to 1,\\[6pt]
		e^{\frac{i m/\kappa}{2u+2}}, & u\to -1.
	\end{cases}
	\label{eq:boundary_u}
\end{equation}
To accommodate the above boundary conditions, we define
\begin{equation}
	\widetilde{\varphi}(u)=e^{-\frac{i m/\kappa}{2u-2}}\,e^{\frac{i m/\kappa}{2u+2}}\,\psi(u).
	\label{eq:modelA_factor}
\end{equation}
Substituting Eq.~(\ref{eq:modelA_factor}) into Eq.~(\ref{eq:sch_u_raw}), we obtain the standard AIM equation
\begin{equation}
	\psi''(u)
	=
	\lambda_0(u)\psi'(u)+s_0(u)\psi(u),
	\label{eq:modelA_aim_standard}
\end{equation}
where
\begin{equation}
	\lambda_0(u)
	=
	-\frac{
		2u\left(
		-3+2u^2+u^4
		+2i(1+u^2)\dfrac{m}{\kappa}
		\right)
	}{
		(u^2-1)^2(1+u^2)
	},
	\label{eq:modelA_lambda0}
\end{equation}

\begin{equation}
	\begin{aligned}
		s_0(u)
		={}&
		\frac{1}{
			4(u^2-1)^2(u^2+1)
			(u^4-u^2+1)^2
			(5u^4-8u^2+5)^2
		}
		\\[4pt]
		&\times
		\Bigg[
		3(u^2+1)^3
		\left(
		70-311u^2+642u^4-798u^6
		+642u^8-311u^{10}+70u^{12}
		\right)
		\\[4pt]
		&\qquad
		+8i(u^2-1)
		(u^4-u^2+1)^2
		(5u^4-8u^2+5)^2
		\frac{m}{\kappa}
		\\[4pt]
		&\qquad
		-4(u^2+1)
		(u^4-u^2+1)^2
		(5u^4-8u^2+5)^2
		\frac{m^2}{\kappa^2}
		\Bigg].
	\end{aligned}
	\label{eq:s0_u_simplified}
\end{equation}

\begin{figure}[htbp]
	\centering
	\includegraphics[width=0.48\textwidth]{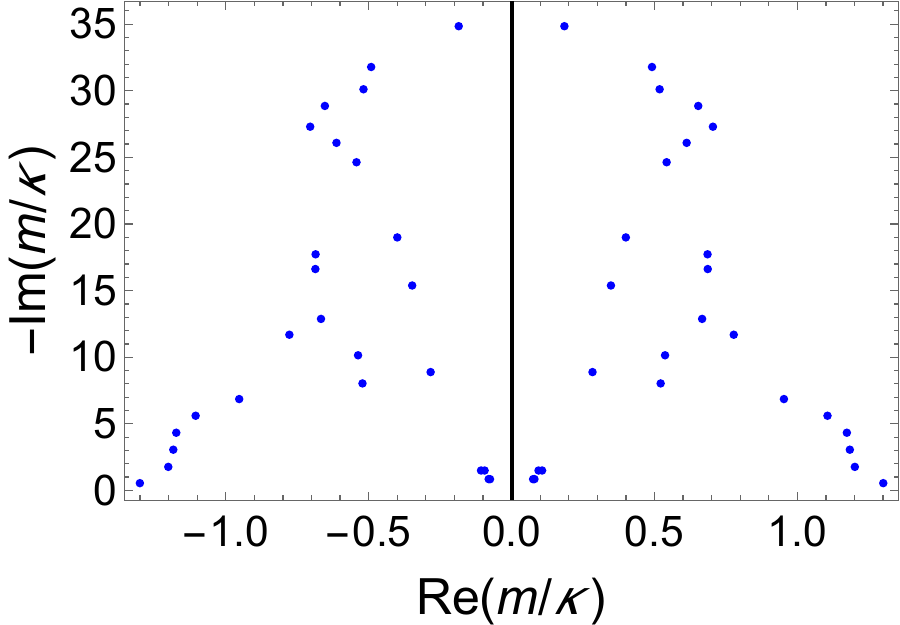}
	\caption{The QNFs of Model A obtained using the AIM in the complex $m/\kappa$ plane
		with 150 iterations. }
	\label{fig:modelA_aim_qnms}
\end{figure}

Using this quantization condition, we obtain the QNFs of Model A, as shown in Fig.~\ref{fig:modelA_aim_qnms}. The modes shown
have negative imaginary parts, indicating damped behavior.

As an independent frequency-domain approach, we also employ the  Bernstein spectral method to compute the QNMs of the thick brane. Spectral methods provide an efficient framework for solving eigenvalue problems and have been widely used in QNM calculations \cite{Boyd_2001,Trefethen_2000,Jansen_2017,Fortuna_2023}. The Bernstein spectral method employs the same coordinate
compactification and boundary conditions as the AIM; therefore, the
corresponding derivation is not repeated here. The perturbation function is then expanded in a Bernstein polynomial basis on the compact interval, thereby reducing the differential
equation to a generalized matrix eigenvalue problem. Solving the resulting generalized eigenvalue problem gives the QNFs. The spectral discretization is implemented using the Mathematica
package \texttt{SpectralBP}~\cite{Fortuna_2023}. The first six QNFs obtained by the AIM and the Bernstein spectral method are compared in Table~\ref{tab:qnm_modelA_fre}, showing close
agreement between the two methods. 
\begin{table}[htbp]
	\centering
	\caption{The first six quasinormal frequencies of Model A obtained by
		the AIM and the Bernstein spectral method.}
	\label{tab:qnm_modelA_fre}
	
	\small
	\renewcommand{\arraystretch}{1.15}
	\begin{tabular}{ccccc}
		\toprule
		$n$
		& \multicolumn{2}{c}{AIM}
		& \multicolumn{2}{c}{Bernstein spectral method} \\
		\cmidrule(lr){2-5}
		& $\operatorname{Re}(m/\kappa)$
		& $\operatorname{Im}(m/\kappa)$
		& $\operatorname{Re}(m/\kappa)$
		& $\operatorname{Im}(m/\kappa)$ \\
		\midrule
		1 & 1.30060  & -0.546061 & 1.30060  & -0.546061 \\
		2 & 1.20115  & -1.76785  & 1.20105  & -1.76787  \\
		3 & 1.18365  & -3.05646  & 1.18352  & -3.05715  \\
		4 & 1.17323  & -4.33061  & 1.17019  & -4.33000  \\
		5 & 1.10548  & -5.60563  & 1.11033  & -5.59835  \\
		6 & 0.953035 & -6.85803  & 0.925606 & -6.89123  \\
		\bottomrule
	\end{tabular}
\end{table}

\subsubsection{Time evolution of Gaussian wave packets}
The frequency-domain calculations determine the QNFs but do not
directly describe the temporal evolution of the perturbations. To
obtain the oscillation and decay of the waveforms, we evolve Gaussian
wave packets in the time-domain and extract the dominant QNFs from the
resulting signals~\cite{Tan_2024,Tan_2025_graviscalar}. For the
numerical evolution, we introduce the coordinates
\(u=t-z\) and \(v=t+z\). In terms of these coordinates,
Eq.~\eqref{eq:timewave} becomes
\begin{equation}
\left( \frac{4 \partial^2}{\partial u \partial v} + V_g + p^2 \right) \Phi(u,v) = 0.
\end{equation}

\begin{figure*}[htb]
	\centering
	\begin{subfigure}[b]{0.42\textwidth}
		\centering
		\includegraphics[width=\linewidth]{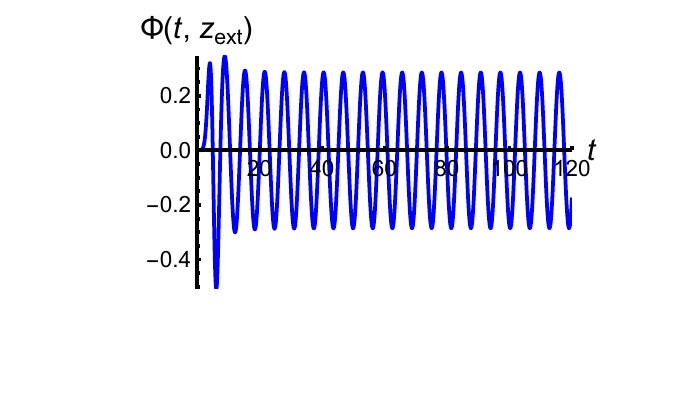}
		\caption{$\kappa z_{\rm ext}=0$}
		\label{fig:Gauss_linear_z0}
	\end{subfigure}
\hspace{0.06\textwidth}
	\begin{subfigure}[b]{0.42\textwidth}
		\centering
		\includegraphics[width=\linewidth]{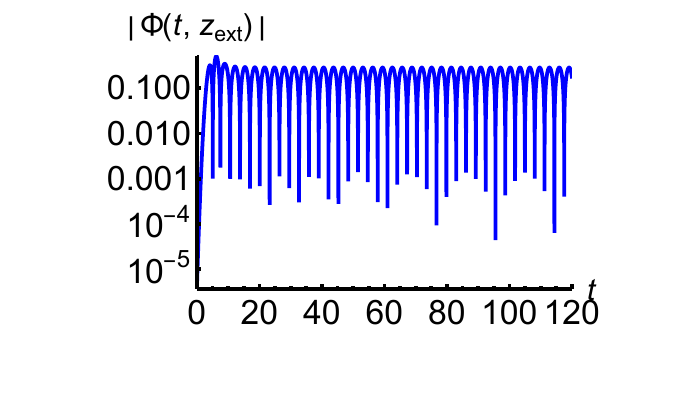}
		\caption{$\kappa z_{\rm ext}=0$}
		\label{fig:Gauss_log_z0}
	\end{subfigure}
	
	\medskip
	\begin{subfigure}[b]{0.42\textwidth}
		\centering
		\includegraphics[width=\linewidth]{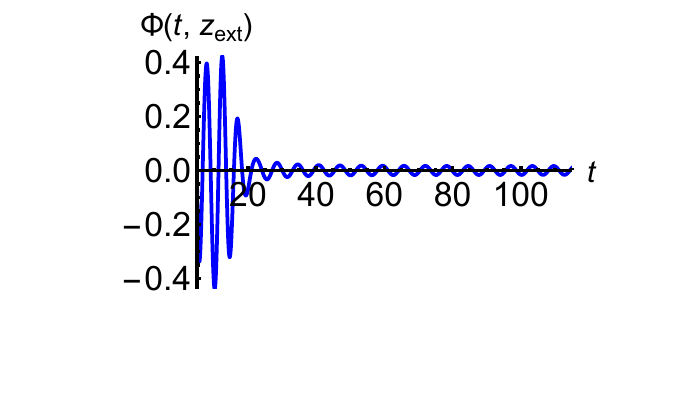}
		\caption{$\kappa z_{\rm ext}=5$}
		\label{fig:Gauss_linear_z5}
	\end{subfigure}
   \hspace{0.06\textwidth}
	\begin{subfigure}[b]{0.42\textwidth}
		\centering
		\includegraphics[width=\linewidth]{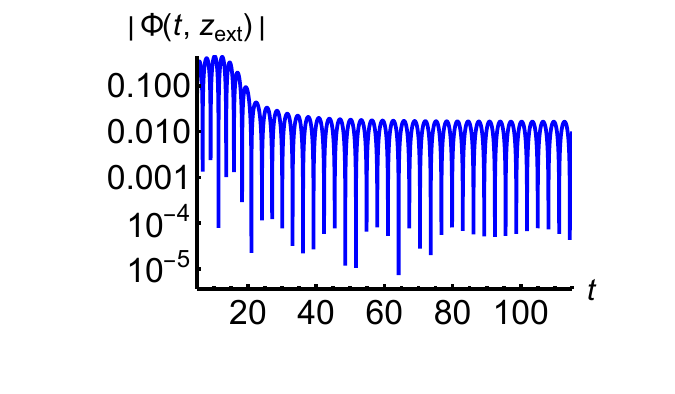}
		\caption{$\kappa z_{\rm ext}=5$}
		\label{fig:Gauss_log_z5}
	\end{subfigure}
	
	\medskip
	\begin{subfigure}[b]{0.42\textwidth}
		\centering
		\includegraphics[width=\linewidth]{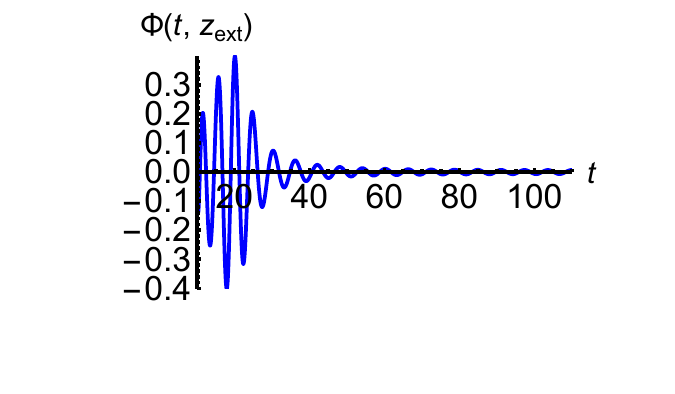}
		\caption{$\kappa z_{\rm ext}=10$}
		\label{fig:Gauss_linear_z10}
	\end{subfigure}
\hspace{0.06\textwidth}
	\begin{subfigure}[b]{0.42\textwidth}
		\centering
		\includegraphics[width=\linewidth]{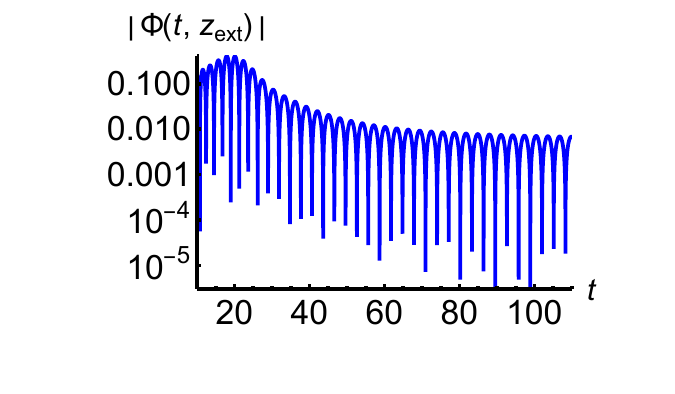}
		\caption{$\kappa z_{\rm ext}=10$}
		\label{fig:Gauss_log_z10}
	\end{subfigure}
	
	\caption{Left panels: Time evolution of the Gaussian pulse at different
		extraction points. The signals are extracted at $\kappa z_{\rm ext}=0$, $5$,
		and $10$. Right panels: The corresponding results shown on a logarithmic
		scale.}
	\label{fig:Gauss}
\end{figure*}

For the time-domain analysis, we first consider a Gaussian initial wave packet. The wave function is given by
\begin{equation}
\Phi(0, v) = e^{-\frac{(v - v_c)^2}{2 \sigma^2}}, \quad \Phi(u, 0) = e^{-\frac{v_c^2}{2 \sigma^2}},
\end{equation}
where \( k v_c = 5 \) and \( k \sigma_c = 1 \). This describes a left-moving Gaussian pulse. We further set \( p/\kappa = 1 \). The evolution of the Gaussian pulse is shown in Fig.~\ref{fig:Gauss}. At early times, the waveform is mainly determined by the initial
data and subsequently evolves into a plane wave, whose frequency
and maximum amplitude remain constant in time. From Figs.~\ref{fig:Gauss_linear_z0},
\ref{fig:Gauss_linear_z5}, and
\ref{fig:Gauss_linear_z10}, we can see that the oscillation
frequency of the late-time signal is approximately independent
of the extraction point. In contrast,
Figs.~\ref{fig:Gauss_log_z0},
\ref{fig:Gauss_log_z5}, and
\ref{fig:Gauss_log_z10} show that the late-time amplitude depends
on the extraction position and decreases as the extraction point
moves farther away from the brane. 
Previous time-domain studies have shown that
a Gaussian pulse can excite the localized graviton zero mode,
whose frequency satisfies \(\omega=p\) for \(m=0\)
~\cite{Tan_2022_QNM}. To avoid a zero  mode contribution
and examine the quasinormal ringing more clearly, we next consider
an odd initial wave packet.

\begin{equation}
	\begin{aligned}
		\Phi(0,v)
		&=
		\sin\left(\frac{\kappa v}{2}\right)
		\exp\left(-\frac{\kappa^2v^2}{4}\right), \\
		\Phi(u,0)
		&=
		-\sin\left(\frac{\kappa u}{2}\right)
		\exp\left(-\frac{\kappa^2u^2}{4}\right).
	\end{aligned}
	\label{eq:modelA_odd_initial}
\end{equation}

\begin{figure*}[htb]
	\centering
	
	\begin{subfigure}[b]{0.42\textwidth}
		\centering
		\includegraphics[width=\linewidth]{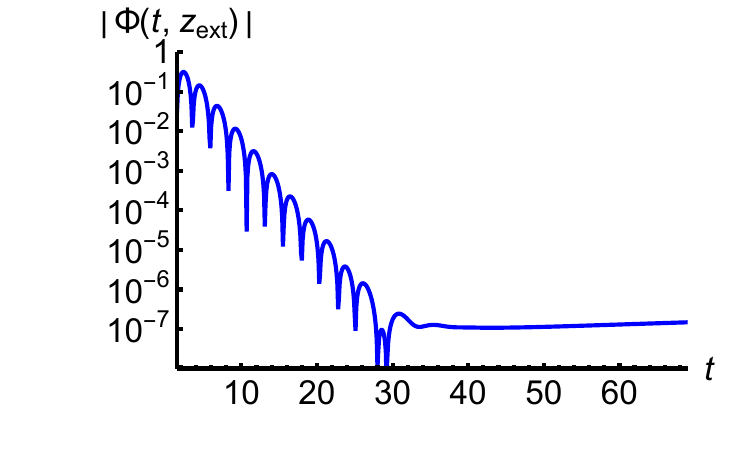}
		\caption{$kz_{\text{ext}}=1$}
		\label{fig:odd_Gauss_a}
	\end{subfigure}
	\hspace{0.06\textwidth}
	\begin{subfigure}[b]{0.42\textwidth}
		\centering
		\includegraphics[width=\linewidth]{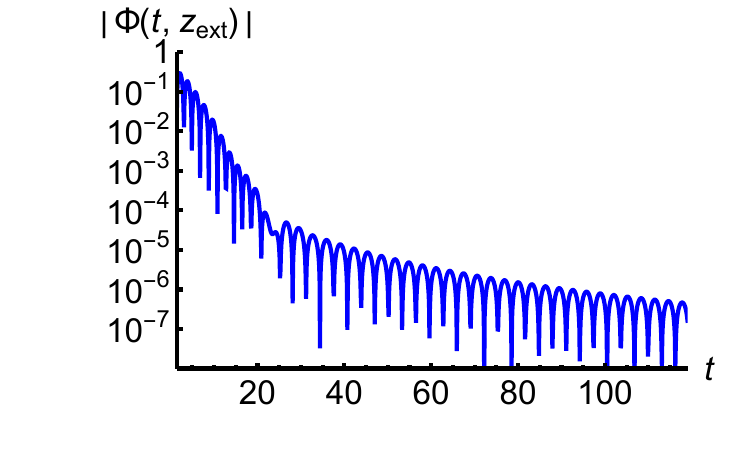}
		\caption{$kz_{\text{ext}}=1$}
		\label{fig:odd_Gauss_b}
	\end{subfigure}
	
	\medskip
	
	\begin{subfigure}[b]{0.42\textwidth}
		\centering
		\includegraphics[width=\linewidth]{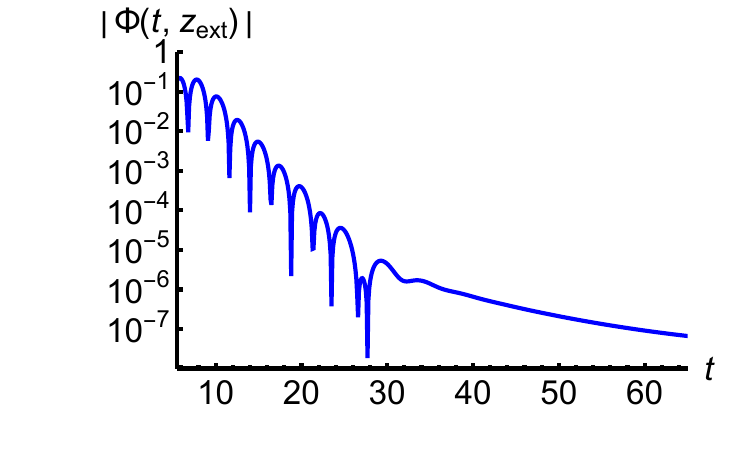}
		\caption{$kz_{\text{ext}}=5$}
		\label{fig:odd_Gauss_c}
	\end{subfigure}
	\hspace{0.06\textwidth}
	\begin{subfigure}[b]{0.42\textwidth}
		\centering
		\includegraphics[width=\linewidth]{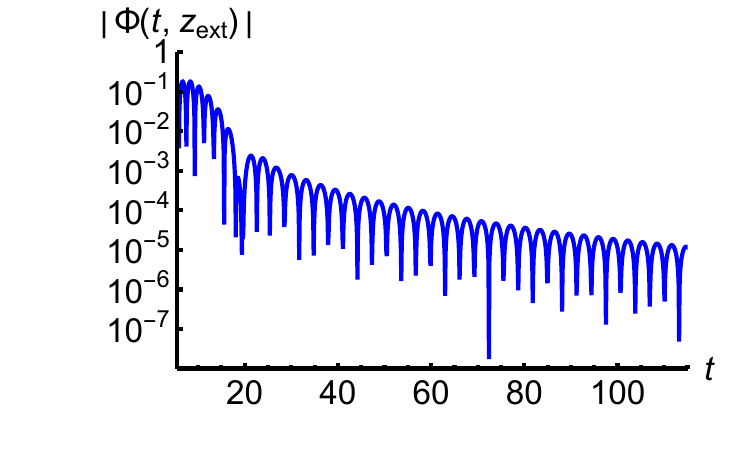}
		\caption{$kz_{\text{ext}}=5$}
		\label{fig:odd_Gauss_d}
	\end{subfigure}
	
	\medskip
	
	\begin{subfigure}[b]{0.42\textwidth}
		\centering
		\includegraphics[width=\linewidth]{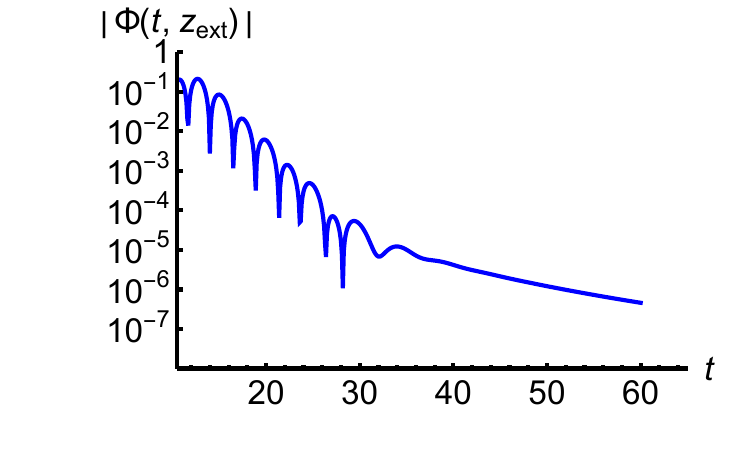}
		\caption{$kz_{\text{ext}}=10$}
		\label{fig:odd_Gauss_e}
	\end{subfigure}
	\hspace{0.06\textwidth}
	\begin{subfigure}[b]{0.42\textwidth}
		\centering
		\includegraphics[width=\linewidth]{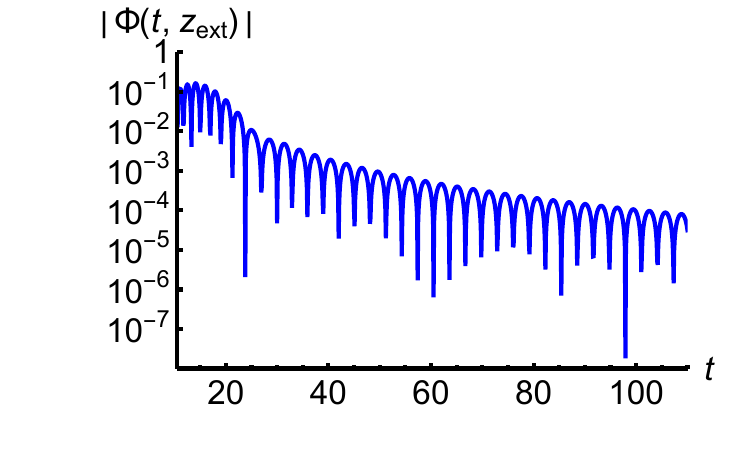}
		\caption{$kz_{\text{ext}}=10$}
		\label{fig:odd_Gauss_f}
	\end{subfigure}
	
	\caption{Left panels: Time evolution of the odd wave packet at different
		extraction points for $p=0$. Right panels: Time evolution of the odd wave
		packet at different extraction points for $p=1$.}
	\label{fig:odd_Gauss}
\end{figure*}

Figure \ref{fig:odd_Gauss} shows the evolution of the odd wave packet at different extraction points for two cases, \(p/\kappa = 0\) and \(p/\kappa = 1\). For $p/\kappa=0$, the evolution can be divided into three stages.
(a) In the initial burst stage, the waveform is mainly determined
by the chosen initial perturbation.
(b) During the exponential decay stage, the oscillation frequency
and damping time are determined by the characteristic structure
of the thick brane and are independent of the specific initial
configuration used to excite the perturbation.
(c) At sufficiently late times, the signal enters a power-law
decay stage, similar to the late-time behavior of a massless field
around a Schwarzschild black hole.  Since the first QNM has
the smallest damping rate and hence the longest lifetime, it
dominates the ringdown stage. We therefore determine its complex
frequency by fitting the ringdown interval, excluding the initial
burst and the late-time tail, with the damped sinusoidal form
\begin{equation}
	\Phi(t,z_{\rm ext})
	\simeq
	A e^{-|m_I|t}
	\cos\left(m_R t+\delta\right),
	\label{eq:modelA_ringdown_fit}
\end{equation}
where \(m_R\) and \(m_I\) denote the real and imaginary parts of
the QNF, respectively, and the magnitude of \(m_I\) characterizes
the damping rate. Applying this fit to the \(p=0\) ringdown signal extracted at
\(z_{\rm ext}=1\), shown in Fig.~\ref{fig:odd_Gauss_a}, we obtain
\(m/\kappa=1.30501-0.536335i\).
This result is in good agreement with those obtained using the AIM
and the Bernstein spectral method.
For \( p/\kappa = 1 \), the waveform also exhibits damped oscillatory behavior.

Overall, the time evolution confirms the presence of damped QNMs in
Model A. The fundamental frequency extracted from the ringdown is
consistent with the results obtained by the AIM and the Bernstein
spectral method. These modes provide dynamical information about the
brane beyond that carried by the localized graviton zero mode.

\subsection{Quasinormal modes of Model B}

\label{subsec:qnm_modelB}
As shown in  Sec.~\ref{subsec:modelB_localization}, Model B supports long-lived resonant KK
 modes. In this subsection,  we investigate the QNMs of Model B and examine their relation to the KK
resonances. We employ two independent numerical approaches: the direct integration method in the
frequency domain and the time evolution method~\cite{Pani_2013,Tan_2024,Zhang2026}. Since the
time-domain scheme has already been introduced in Sec.~\ref{subsec:qnm_modelA}, here we only briefly summarize the numerical procedure. Throughout this subsection, we set the momentum along the brane to
\(p=0\).

In the frequency domain, we calculate the QNFs of Model B using
the direct integration method. The basic idea is to integrate the
Schr\"odinger-like equation numerically from the outer boundary
toward the brane center for a trial complex value of \(m\), while
imposing the appropriate boundary conditions. Since the effective potential of the thick brane is symmetric about $z=0$, the mode functions can be classified according to their parity as even and
odd modes.
Accordingly, a purely outgoing wave condition is imposed at a
sufficiently large positive \(z\), whereas the corresponding parity
condition is imposed at the brane center. The values of
\(m\) satisfying these conditions determine the QNFs of Model B.

\begin{figure*}[htbp]
	\centering
	
	\begin{subfigure}[b]{0.42\textwidth}
		\centering
		\includegraphics[width=\linewidth]
		{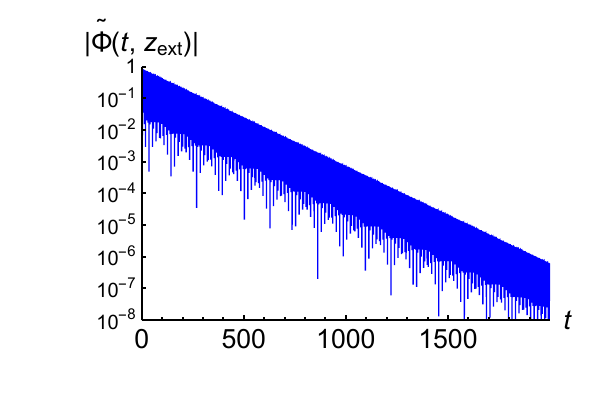}
		\caption{\(k=1,\ n=1\)}
		\label{fig:modelB_qnm_k1_n1}
	\end{subfigure}
	\hspace{0.06\textwidth}
	\begin{subfigure}[b]{0.42\textwidth}
		\centering
		\includegraphics[width=\linewidth]
		{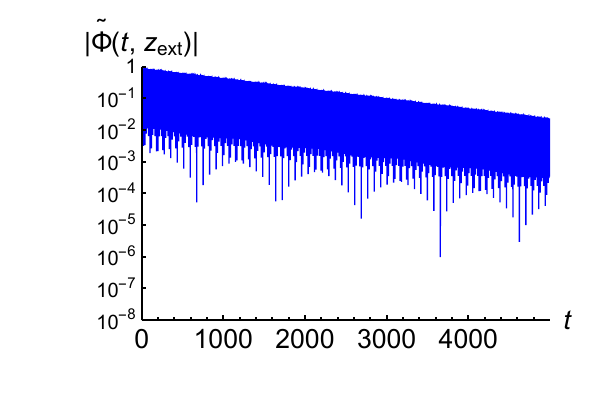}
		\caption{\(k=3,\ n=1\)}
		\label{fig:modelB_qnm_k3_n1}
	\end{subfigure}
	
	\medskip
	
	\begin{subfigure}[b]{0.42\textwidth}
		\centering
		\includegraphics[width=\linewidth]
		{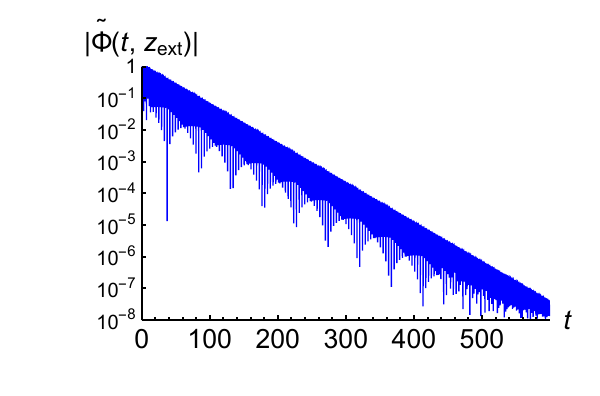}
		\caption{\(k=3,\ n=2\)}
		\label{fig:modelB_qnm_k3_n2}
	\end{subfigure}
	\hspace{0.06\textwidth}
	\begin{subfigure}[b]{0.42\textwidth}
		\centering
		\includegraphics[width=\linewidth]
		{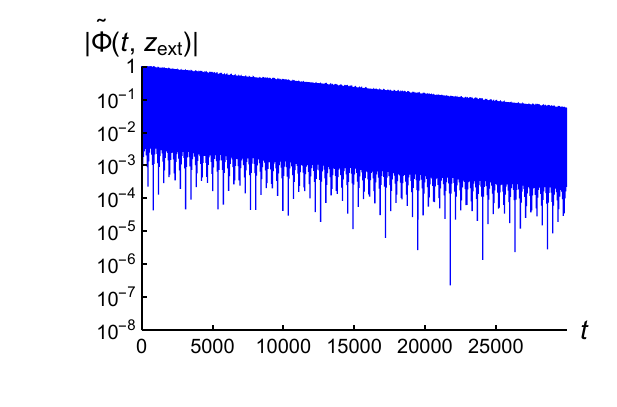}
		\caption{\(k=15,\ n=1\)}
		\label{fig:modelB_qnm_k15_n1}
	\end{subfigure}
	
	\medskip
	
	\begin{subfigure}[b]{0.42\textwidth}
		\centering
		\includegraphics[width=\linewidth]
		{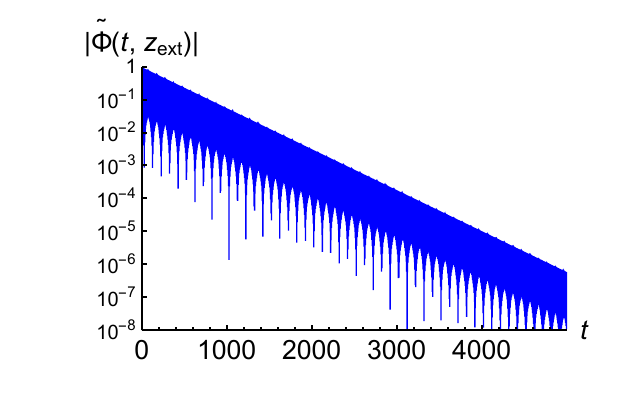}
		\caption{\(k=15,\ n=2\)}
		\label{fig:modelB_qnm_k15_n2}
	\end{subfigure}
	\hspace{0.06\textwidth}
	\begin{subfigure}[b]{0.42\textwidth}
		\centering
		\includegraphics[width=\linewidth]
		{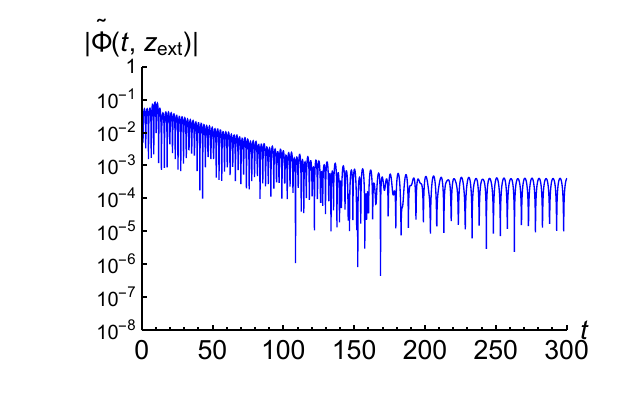}
		\caption{\(k=15,\ n=3\)}
		\label{fig:modelB_qnm_k15_n3}
	\end{subfigure}
	
	\caption{
		Time-domain waveforms associated with the QNMs of Model B
		for \(\xi=0.99\). All perturbation amplitudes are extracted
		at the common observation point \(z_{\rm ext}=1\) and plotted
		on a logarithmic scale.
	}
	\label{fig:modelB_qnm_time_domain_all}
\end{figure*}

As an independent check of the frequency-domain results, we also
perform a time-domain calculation for Model B. 
For each mode, the corresponding resonant
wave function of the dual potential is taken as the initial profile,
\begin{equation}
	\widetilde{\Phi}(0,z)=\widetilde{\varphi}_n(z),
	\qquad
	\partial_t\widetilde{\Phi}(0,z)=0.
\end{equation}
The QNFs are extracted from the ringdown signals using the same damped
sinusoidal fitting procedure as in Eq.~\eqref{eq:modelA_ringdown_fit}.
Figure~\ref{fig:modelB_qnm_time_domain_all} shows the time-domain waveforms associated with the QNMs of Model B on a logarithmic amplitude scale.
The waveform is recorded at \(z_{\rm ext}=1\), where
\(|\widetilde{\Phi}(t,z_{\rm ext})|\) denotes the absolute amplitude
of the time-dependent perturbation evolved under the dual potential.
After the initial transient, the signals exhibit damped oscillations
with approximately exponential decay, similar to the behavior found
for \(f(R)\) thick branes~\cite{YuPeng_2025_fRQNM}. The QNFs extracted from the time-domain ringdown signals, together with those obtained by the direct integration method, are
summarized in Table~\ref{tab:modelB_qnm_comparison}.  The two methods show good
agreement in both the real and imaginary parts of the QNFs for all
the modes considered. For the same $n$, the real part of the QNF decreases as $k$ increases. Similarly, for the modes available at different values of $k$, the magnitude
of the imaginary part decreases as $k$ increases, indicating weaker damping
and a longer lifetime.
\begin{table}[htbp]
	\centering
	\caption{Comparison of the QNFs obtained from the time evolution and the direct integration method for Model B with $\xi=0.99$.}
	\label{tab:modelB_qnm_comparison}
	\begin{tabular}{c c c c c c}
		\hline
		\multirow{2}{*}{$k$} & \multirow{2}{*}{$n$} 
		& \multicolumn{2}{c}{Time-domain method} 
		& \multicolumn{2}{c}{Direct integration method} \\
		\cline{3-6}
		& & $\mathrm{Re}(m)$ & $\mathrm{Im}(m)$ 
		& $\mathrm{Re}(m)$ & $\mathrm{Im}(m)$ \\
		\hline
		1  & 1 & 1.27729 & $-0.00710$   & 1.27727 & $-0.00706$   \\
		\hline
		\multirow{2}{*}{3} 
		& 1 & 0.91188 & $-0.000732$ & 0.91190 & $-0.000743$ \\
		& 2 & 1.55398 & $-0.028940$  & 1.55368 & $-0.02837$  \\
		\hline
		\multirow{3}{*}{15} 
		& 1 & 0.63294 & $-0.0000971$ & 0.63296 & $-0.0000960$ \\
		& 2 & 1.19770 & $-0.0028649$  & 1.19772 & $-0.002877$  \\
		& 3 & 1.62761 & $-0.032897$   & 1.62752 & $-0.03424$   \\
		\hline
	\end{tabular}
\end{table}

\begin{figure}[htbp]
	\centering
	\begin{subfigure}[b]{0.42\textwidth}
		\centering
		\includegraphics[width=\textwidth]{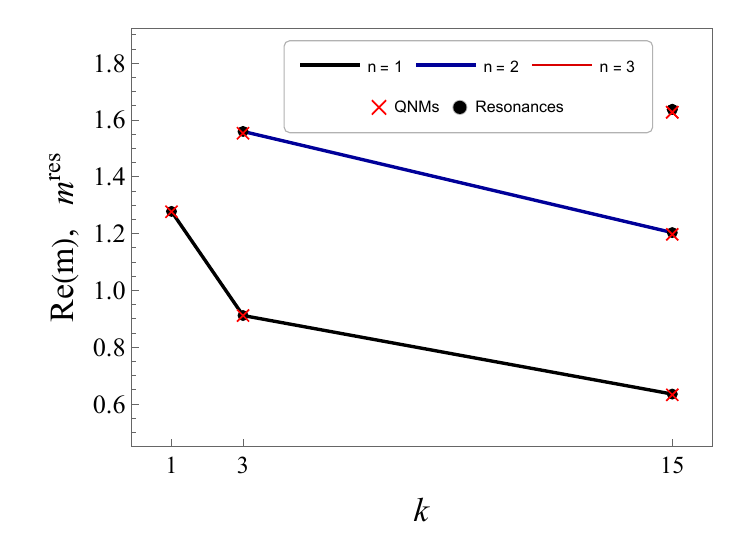}
		\label{fig:modelB_mass}
	\end{subfigure}
\hspace{0.06\textwidth}
	\begin{subfigure}[b]{0.42\textwidth}
		\centering
		\includegraphics[width=\textwidth]{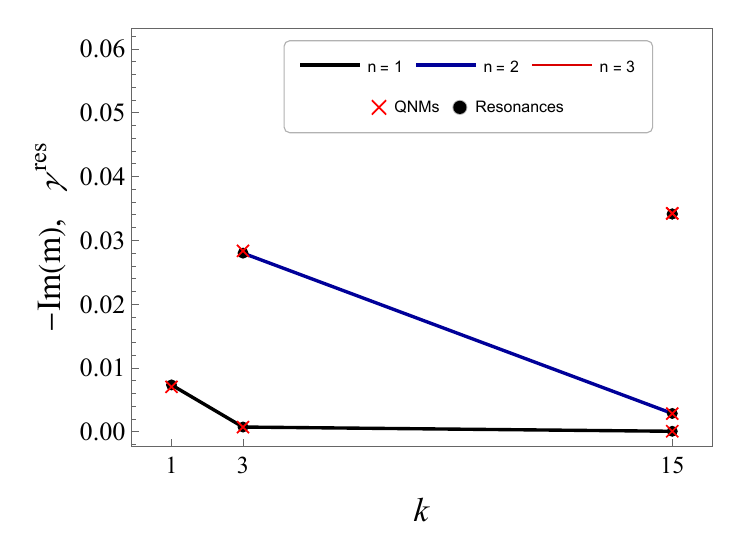}
		\label{fig:modelB_decay}
	\end{subfigure}
	 \caption{Left panel: The KK resonance masses \(m^{\rm res}\) and the real parts
	 	\(\operatorname{Re}(m)\) of the corresponding QNFs obtained by the
	 	direct integration method for different values of \(k\).
	 	Right panel: The decay rates \(\gamma^{\rm res}\) of the KK resonances
	 	and the magnitudes of the imaginary parts of the corresponding QNFs.}
	\label{fig:modelB_qnm_resonance}
\end{figure}

Since both resonant KK modes and QNMs characterize the dynamics of the
thick brane, it is natural to examine their relation. 
Figure~\ref{fig:modelB_qnm_resonance} compares the resonance masses and decay rates with the real parts and the magnitudes of the imaginary parts of the corresponding QNFs, respectively. The two sets of results are in close agreement. 
The agreement arises because the resonant modes and the QNMs are governed by the same Schr\"odinger-like equation and exhibit outgoing-wave behavior in the asymptotic region.
Therefore, the KK resonances and the corresponding QNMs provide
complementary descriptions of the same quasi-localized massive gravitational excitations~\cite{Tan_2024}.

\section{Conclusions and discussion}\label{sec:conclusion}

In this work, we have investigated the characteristic modes of tensor
perturbations in thick branes with a non-minimally coupled bulk scalar field.
We have considered two representative thick brane configurations: a
single-kink thick brane (Model A) and a deformed thick brane with internal
structure (Model B). By reducing the tensor perturbation equation to a
Schr\"odinger-like equation, we study the localization of the graviton zero
mode, the resonant KK modes, and the QNMs. The relation between resonant KK modes and QNMs is further examined, showing that they provide complementary descriptions of the same quasi-localized massive gravitational excitations in non-minimally coupled thick branes.

For Model A, we focus on the special case $\xi=1/7$, for which the tensor effective potential can be obtained analytically. The potential has a
volcano-like structure and approaches zero in the asymptotic region. The
graviton zero mode is normalizable and localized on the brane, as shown
in Fig.~\ref{fig:modelA_Vg_zero}. However, no pronounced resonant KK modes are found for this parameter choice.
For Model B, the non-minimal coupling parameter $\xi$ and the deformation
parameter $k$ significantly modify the internal structure of the brane and the
corresponding effective potential. As illustrated in Fig.~\ref{fig:modelB_vgz_k}, a larger $\xi$ enhances the double-barrier structure, whereas the separation between the barriers grows with $k$. The graviton zero mode remains localized for all the parameter
values considered. For smaller values of $\xi$, only
a few resonant KK modes are found, or none at all. We therefore focus on
$\xi=0.99$, where the double-barrier structure is most pronounced. In this regime, the relative probability spectra shown in
Fig.~\ref{fig:relative_probability_k} reveal an increase in the number of
resonant KK modes from one at $k=1$ to three at $k=15$. Moreover, the first
resonant mode becomes progressively longer lived as the brane deformation
becomes more pronounced, indicating enhanced trapping of massive gravitational
excitations.

We further calculate the QNMs of both brane models using independent
frequency-domain and time-domain approaches. For Model A, the QNFs obtained using the AIM and the Bernstein spectral
method agree closely for the low overtones, as summarized in
Table~\ref{tab:qnm_modelA_fre}. The fundamental mode extracted from the time-domain
ringdown shown in Fig.~\ref{fig:odd_Gauss} is also consistent with
the frequency-domain results. For Model B, the QNFs extracted from the time-domain ringdown are consistent with those obtained by the direct integration method, as
summarized in Table~\ref{tab:modelB_qnm_comparison}. As the
deformation parameter $k$ increases, the real part of the QNF decreases while
the magnitude of the imaginary part becomes smaller, corresponding to lower
oscillation frequencies and weaker damping. The agreement between the two methods
confirms the reliability of the QNM extraction.

Finally, we compare the resonant KK modes with the corresponding QNMs in
Model B. As shown in Fig.~\ref{fig:modelB_qnm_resonance}, the resonance masses agree well with
the real parts of the corresponding QNFs, while the decay rates obtained
from the time evolution of the resonant states are consistent with the
magnitudes of the imaginary parts. These results indicate that resonant KK
modes and QNMs provide complementary descriptions of the same quasi-localized
massive gravitational excitations in non-minimally coupled thick branes.

Several directions remain for future investigation. Motivated by studies of
quasinormal ringing in de Sitter braneworlds~\cite{Jia_2025_dS}, it would be
interesting to extend the present analysis to non-minimally coupled curved
thick branes, such as de Sitter and anti-de Sitter branes. Another direction is to investigate
the contributions of massive KK modes to the 4D Newtonian potential.
In addition, possible gravitational echoes associated with the double-barrier
structure of the effective potential in the deformed brane may be explored.

\acknowledgments
We would like to thank Prof. Qin Tan for helpful discussions on numerical methods for extracting
quasinormal modes and for valuable comments and suggestions that helped improve
the manuscript. This work is supported by the National Natural Science Foundation of China (Grants
No. 11305119), the Natural Science Foundation of Shaanxi Province (No. 2022JQ-037), and
the 111 Project (B17035).

\bibliographystyle{JHEP}

\end{document}